\documentclass[
reprint,
superscriptaddress,
longbibliography,
bibnotes,
amsmath,amssymb,
aps,
prb,
]{revtex4-2}
\usepackage[utf8]{inputenc}
\usepackage[english]{babel}
\usepackage[plainpages = false, pdfpagelabels, 
                 bookmarks,
                 bookmarksopen = true,
                 bookmarksnumbered = true,
                 breaklinks = true,
                 linktocpage,
                 colorlinks = true,
                 linkcolor = blue,
                 urlcolor  = blue,
                 citecolor = blue,
                 anchorcolor = green,
                 hyperindex = true,
                 hyperfigures
                 ]{hyperref} 
\usepackage{enumerate}
\usepackage{tabularx}
\usepackage{makecell,tabularx}

\usepackage{multirow}
\usepackage{booktabs}
\usepackage{makecell, multirow}
\usepackage{tikz}
\usetikzlibrary{patterns}
\usepackage{slashed}
\usepackage{physics}
\usepackage{verbatim}
\usepackage{cancel}
\usepackage{bbold}
\usepackage{bm}
\usepackage{booktabs}
\usepackage[caption=false,singlelinecheck=off,labelformat=simple]{subfig}

\usepackage{blindtext}
\usepackage{bbm}
\usepackage{graphicx}
\usepackage{dcolumn}
\usepackage{bm}
\usepackage[normalem]{ulem}

\usepackage{xcolor}
\definecolor{MyViolet}{RGB}{153,  0,153}
\definecolor{FloBlue}{rgb}{0.25, 0.41, 0.88}

\begin{document}
\title{Impact of potential and temperature fluctuations on charge and heat transport in quantum Hall edges in the heat Coulomb blockade regime}
\author{Christian Sp\r{a}nsl\"{a}tt}
\thanks{These authors contributed equally to this work.}
\affiliation{Department of Microtechnology and Nanoscience (MC2),Chalmers University of Technology, S-412 96 G\"oteborg, Sweden}
\author{Florian St\"abler}
\thanks{These authors contributed equally to this work.}
\affiliation{D\'epartement de Physique Th\'eorique, Universit\'e de Gen\`eve, CH-1211 Gen\`eve 4, Switzerland}
\author{Eugene V. Sukhorukov}
\affiliation{D\'epartement de Physique Th\'eorique, Universit\'e de Gen\`eve, CH-1211 Gen\`eve 4, Switzerland}
\author{Janine Splettstoesser}
\affiliation{Department of Microtechnology and Nanoscience (MC2),Chalmers University of Technology, S-412 96 G\"oteborg, Sweden}

\begin{abstract}
We present a broad study of charge and heat transport in a mesoscopic system where one or several quantum Hall edge channels are strongly coupled to a floating Ohmic contact (OC). 
It is well known that charge-current fluctuations emanating from the OC along the edge channels are highly susceptible to the OC charge capacitance in the heat Coulomb blockade regime (an impeded ability of the OC to equilibrate edge channels). Here, we demonstrate how potential- and temperature fluctuations due to finite OC charge and heat capacities impact the heat-current fluctuations emitted from the OC.
First, by assuming an infinite OC heat capacity, we show that the output heat-current noise is strongly dependent on the OC charge capacitance, following from a close relation between one-dimensional charge- and heat currents. When also the OC heat capacity is finite, an interplay of potential- and temperature fluctuations influences the heat transport. Concretely, we find that the effect of the charge capacitance on heat transport manifests in terms of a strongly increased energy relaxation time in the heat Coulomb blockade regime. Furthermore, we find expressions for a broad set of output observables, such as charge and heat auto- and cross correlations, as functions of input and OC fluctuations, depending on the relation between charge and energy relaxation times compared to the frequency of fluctuations and inverse (local) temperatures as well as on the number of edge channels attached to the OC. 
Finally, we show that a finite OC heat capacity transforms the full counting statistics of the output charge from Gaussian to non-Gaussian. Our findings provide novel opportunities to experimentally probe and harness the quantum nature of heat transport in strongly coupled electron circuits. 
\end{abstract}

\date{\today}
\maketitle
\section{Introduction} 
Manipulation and detection of heat currents in nanoscale circuits have in recent years enabled detailed experimental investigations of the quantum nature of heat conduction~\cite{Pekola21Oct}. These advancements provide plenty of opportunities for both fundamental discovery, e.g., quantum bounds on heat and information transfer~\cite{Pendry1983Jul}, as well as for developing novel quantum technologies, such as nanoscale heat-to-work conversion~\cite{Benenti2017Jun,Whitney2019Apr}. 

\begin{figure}[t!]
\includegraphics[trim={1mm 0 0 0},clip,width =0.99\columnwidth]{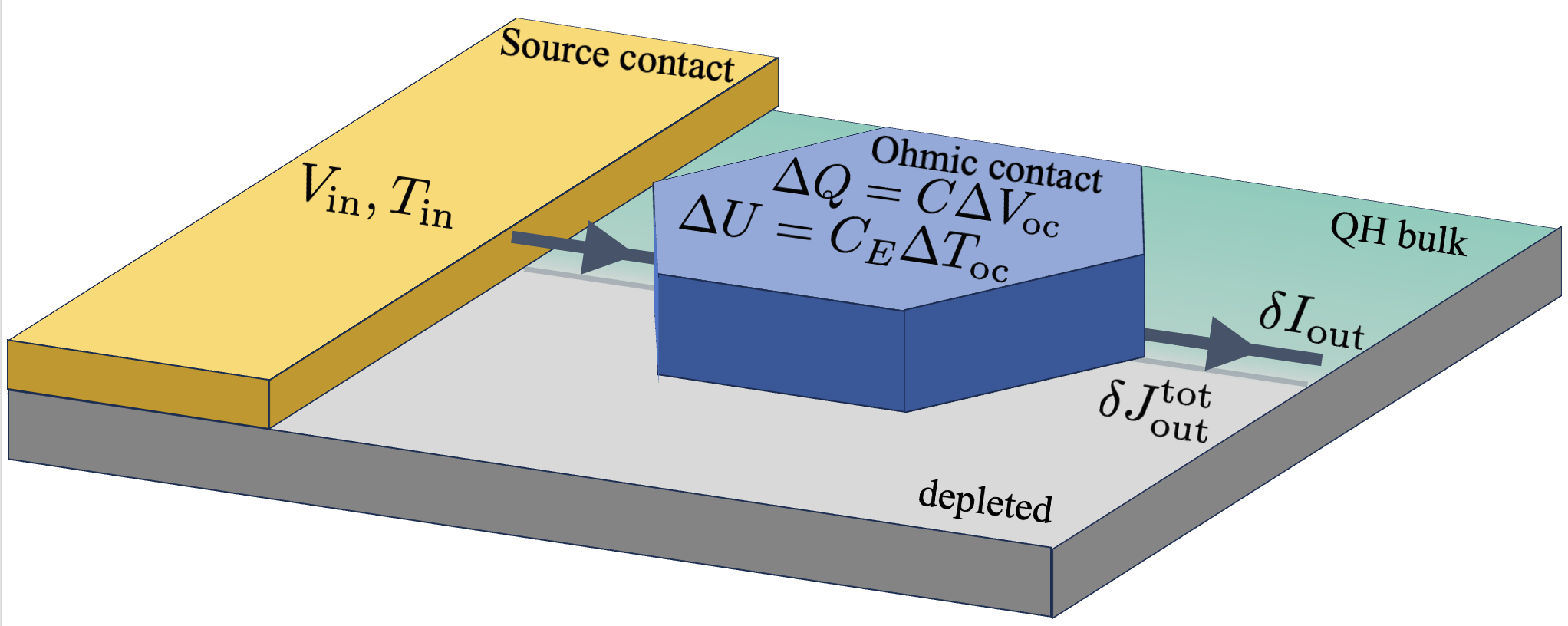}
\caption{Schematics of a floating Ohmic contact (OC, blue) connected to a single chiral, one-dimensional quantum Hall edge channel (dark gray arrow). The OC sits on the boundary (black, dotted line) between the quantum Hall bulk region (turquoise) and a fully depleted region (light gray). The edge channel emanates from a large metallic contact (yellow), characterized by the voltage $V_{\rm in}$ and the temperature $T_{\rm in}$. Fluctuations carried by the edge channel induce potential- and temperature fluctuations,  $\Delta V_{\rm oc}$ respectively $\Delta T_{\rm oc}$, in the OC. In turn, these quantities produce charge-current $\delta I_{\rm out}$ and heat-current $\delta J_{\rm out}^\mathrm{tot}$ fluctuations in the outgoing channel.
}
\label{fig:Single_Channel_Setup}
\end{figure}

A versatile platform for this scope in the context of electron transport, is a small, floating Ohmic contact (OC) coupled to one or several one-dimensional (1D) chiral, ballistic quantum Hall (QH) edge channels~\cite{Halperin1982Feb,Buttiker1988Nov,Wen1990Jun}, see Fig.~\ref{fig:Single_Channel_Setup}. 
Such a floating ``probe''~\cite{Buttiker1988Jan,Buttiker1988Nov} has previously been used to study a wide range of fascinating quantum phenomena, both theoretically and experimentally, such as equilibrium and and out-of-equilibrium dynamical Coulomb blockade \cite{duprez_dynamical_2021,altimiras_dynamical_2014}, heat Coulomb blockade~\cite{Sivre2018Feb,sivre_electronic_2019}, electron state teleportation~\cite{Idrisov2018Jul,Duprez2019}, Luttinger liquid behaviour~\citep{anthore_circuit_2018,jezouin_tomonagaluttinger_2013,anthore_universality_2020}, charge fractionalization~\cite{idrisov_quantum_2020,morel_fractionalization_2022}, Coulomb mediated heat drag~\cite{Idrisov2022Sep}, or tunable multi-channel Kondo effects~\cite{Iftikhar2015Oct,Iftikhar2018Jun}. It has also been proposed as a probe for temperature fluctuations~\cite{Battista2013Mar,Utsumi2014May,vandenBerg2015Jul,Dashti2018Aug}.

At the heart of these studies is a key parameter of the single edge channel-OC system, namely the $RC$ time, $\tau_C\equiv R_q C$, where $C$ is the total OC charge capacitance and $R_q\equiv 2\pi/e^2$ is the resistance quantum (in units where the reduced Planck constant $\hbar=1$). The $RC$ time is the characteristic time scale on which the OC responds to fluctuations of the charge and it is closely related to the OC charging energy as $E_C \equiv e^2/(2C)=\pi\tau_C^{-1}$. At low temperatures, the combination of chiral electron transport and a sufficiently short $RC$ time leads to the OC acting as a highly sensitive band-pass filter for the edge channels. 

Whereas most previous works on the edge channel-OC system focused on transport of average charge and heat currents, as well as on charge-current fluctuations, far less understood are the heat-current fluctuations~\cite{Pekola21Oct, Krive2001Nov,Sergi2011Jan,Averin2010Jun,Battista2014Aug,Utsumi2014May,vandenBerg2015Jul,Dashti2018Aug,Crepieux2021Jan,Eriksson2021Sep, Ebisu2022May,Tesser2022Oct,Idrisov2022Sep} emitted from the OC. Heat-current fluctuations play a key role for, e.g., the efficiency and precision in nanoscale heat transport and thermoelectric conversion~\cite{Benenti2017Jun,Tran2023Jan}. In order to understand how this fundamentally interesting system, composed of edge channels and OC, performs with respect to fluctuations of heat transport in the quantum regime, detailed studies are needed.

In this paper, we meet this demand and present an extensive theoretical analysis of heat and charge transport impacted by OC potential- and temperature fluctuations. To achieve this, we determine the combined influence of $\tau_C$ and a second characteristic parameter of the single edge channel-OC system, namely the energy relaxation time $\tau_E\equiv C_E[T_{\rm oc}(t)]/(\kappa_0 T_{\rm oc}(t))$. 
Here, $C_E[T_{\rm oc}(t)]$ is the OC heat capacity [which depends on the temporal OC temperature $T_{\rm oc}(t)$], and $\kappa_0 T_{\rm oc}\equiv(\pi/6) T_{\rm oc}$ is the heat conductance quantum (in units where $\hbar=1$ and the Boltzmann constant $k_{\rm B}=1$). 
Most previous works assumed implicitly  $\omega\tau_E\gg 1$, where $\omega$ is the measurement frequency; a regime which corresponds to the situation where effects from the energy relaxation time are fully neglected, see Ref.~\cite{vandenBerg2015Jul,Dashti2018Aug} for some of the exceptions. 
When considering both OC potential- and temperature fluctuations, but only to leading order, the OC potential fluctuations and the emitted OC charge-current fluctuations are determined solely by the injected charge-current fluctuations and by $\tau_C$. Likewise, the OC temperature fluctuations and the emitted OC heat-current fluctuations are determined only by the injected heat-current fluctuations and by $\tau_E$. 
In other words, the emitted charge- and heat-current fluctuations are fully determined by two separate OC parameters~\cite{vandenBerg2015Jul,Dashti2018Aug}, i.e., $\tau_C$ respectively $\tau_E$. An important question that we  address in the present work is what happens when terms beyond leading order in the OC potential- and temperature fluctuations and their interplay are included in the charge- and heat-current noises.

Our key results are the following: (i) The heat-current fluctuations emanating from the OC are in fact highly influenced by the OC $RC$ time $\tau_C$ even in the limit of infinite OC energy relaxation time, $\tau_E\rightarrow \infty$. This result follows from the close connection between the charge and heat currents carried by 1D chiral edge channels (see Eq.~\eqref{eq:JI_relation} below). The $\tau_C$ dependence in the heat current arises at second order in the OC potential fluctuations. This feature implies that the heat Coulomb blockade effect, i.e.,  a strong suppression of the equilibration between the edge-channel and the OC, is observable not only in charge-current noise measurements~\cite{Sivre2018Feb}, but also in the low-frequency output heat-current noise. 
(ii) The output heat-current noise provides an alternative observable for accessing an effective temperature of the output distribution from the OC. This alternative effective temperature differs from conventional ones based on zero- and finite-frequency charge noise measurements. We show that this discrepancy provides a novel way to study relaxation effects along the edge channels.
(iii) Keeping $\tau_E$ finite (i.e., finite OC heat capacity), we find that both the output heat-current fluctuations and the OC temperature fluctuations are determined by a combination of $\tau_C$ and $\tau_E$. This combination results in similar \textit{band-pass filter effects} (see Eq.~\eqref{eq:SJ_tot_single} below), as the ones that are at the origin of the heat Coulomb blockade (see, e.g., Eq.~\eqref{eq:Sout} below). Moreover, we identify the emergence of a novel energy relaxation time scale, $\tilde{\tau}_E\gg\tau_E$, which depends on the charge relaxation time scale $\tau_C$. This interplay implies that in the heat Coulomb blockade regime, the heat-current fluctuations emanating from the OC become strongly suppressed. The new time scale emerges when we take into account the OC potential fluctuations exactly, and incorporate the effects of OC temperature fluctuations to linear 
order, thereby extending the works in Refs.~\cite{vandenBerg2015Jul,Dashti2018Aug}. 
(iv) For the multi-channel-OC setup, we show that combinations of charge- and heat-current auto- and cross correlations provide straight-forward access to potential- and temperature-fluctuations correlations in the OC, quantities that are highly challenging to extract in a single-channel-OC setup.
(v) A finite energy relaxation time in the multi-channel OC system changes the full counting statistics of the emitted \emph{charges} from Gaussian to non-Gaussian, due to the OC temperature fluctuations. 

The remainder of this paper is organized as follows: In Sec.~\ref{sec:Setup_and_timescales}, we identify a realistic hierarchy of the characteristic OC time scales. This motivates the use of a Langevin approach, which is also presented in that section. In Sec.~\ref{sec:SEC}, we study the single-channel setup in Fig.~\ref{fig:Single_Channel_Setup} and analyze the charge and heat currents and their fluctuations. We also define and discuss effective temperatures of the outgoing, and generally out-of-equilibrium, edge channel. In Sec.~\ref{sec:Multi_Gen}, we extend our analysis to multiple edge channels. Finally, in Sec.~\ref{sec:Non_Gaussian}, we analyze the full-counting statistics of charge transport in presence of temperature fluctuations. Some detailed calculations are delegated to Appendices~\ref{sec:Appendix_A}-\ref{sec:Appendix_equivalence}. Throughout the paper, we use units where $k_{\rm B}=\hbar=h/(2\pi)=1$. 

\section{Setup and formalism}
\subsection{Ohmic contact setup and time scales}
\label{sec:Setup_and_timescales}
We consider the system in Fig.~\ref{fig:Single_Channel_Setup}, consisting of a floating Ohmic contact (OC) connected to a single quantum Hall (QH) edge channel. This channel emanates from another large metallic contact, taken as an ideal, large Fermi reservoir (or ``source'') of electrons. 
The voltage and temperature of this reservoir are $V_{\rm in}$ and $T_{\rm in}$, respectively, both with negligible fluctuations. Unless stated otherwise, we take in this work $V_{\rm in}=0$, which here physically amounts to having no externally applied voltage bias in the device. 
Since the connected edge channel is chiral, it has one incoming and one outgoing segment or ``branch'' with respect to the OC. The charge and heat currents carried by the incoming channel may induce sizeable fluctuations of the OC electrical potential, $V_{\rm oc}(t)\equiv V_{\rm oc}+\Delta V_{\rm oc}(t)$, and the OC temperature, $T_{\rm oc}(t)\equiv T_{\rm oc}+\Delta T_{\rm oc}(t)$, due to the OC's finite charge relaxation time, $\tau_C$, and energy relaxation time, $\tau_E$. 

\begin{table*}[t!]
\centering
\begin{ruledtabular}\begin{tabular}{c c c c c } 
    Time scale &  Typical magnitude  &  Impact & Key equations \\ \hline  
    $\tau_C$ & $10$ - $100$ps~\cite{Sivre2018Feb,Jezouin2013Nov}  & $S_{\rm out}(\omega)$, $J_{\rm out}$, $S^J_{\rm out}(\omega)$ & \eqref{eq:Sout}\eqref{eq:T_integral}\eqref{eq:SQout}  \\
    & & $\kappa$, $\tilde{\tau}_E$, $S^T_{\rm oc}(\omega)$, FCS & \eqref{eq:linear_coefficient}\eqref{eq:New_tau_E}\eqref{eq:ST_equal_T}\eqref{eq:finite_FC_2}\eqref{eq:finite_TC} \vspace{0.25cm} \\
    $\tau_{\rm e-e}$  & $200$ps - $100$ns~\cite{Pothier1997Nov,Katine1998Jan,Pierre2003Aug,Song2013Apr,Song2015Apr} 
     & Time scale of OC thermalisation & \eqref{eq:hierachy_time}  \vspace{0.25cm}\\
    $\tau_E$  & $10$ -$100$$\mu$s~\cite{Jezouin2013Nov}  & $S^{J,\text{tot}}_{\rm out}(\omega)$, $\tilde{\tau}_E$, $S^T_{\rm oc}(\omega)$, FCS  & \eqref{eq:SQtot_cases}\eqref{eq:New_tau_E}\eqref{eq:ST_equal_T}\eqref{eq:finite_FC_2}\eqref{eq:finite_TC}\\
    \\
    \vspace{0.25cm}
    $\tau_{\rm e-ph}$  
    &  $100$$\mu$s at $T \approx 20$mK~\cite{Giazotto2006Mar,Gasparinetti2015Jan}    & Largest time scale, phonons negligible & \eqref{eq:hierachy_time} 
      \\
      \end{tabular}
\end{ruledtabular}
      \caption{Summary of relevant time scales, their impact on the observables treated in this work, and reference to key equations. FCS abbreviates Full Counting Statistics, see Sec.~\ref{sec:Non_Gaussian}. }
    \label{tab:timescales}
\end{table*}

We estimate these relaxation times as follows. The charge relaxation rate is related to the charging energy, $E_C=\pi \tau_C^{-1}$. 
Typically, the charging energy $E_C$, namely the energy required to add an electron to the OC, scales with the OC charge capacitance $C$ via $D/L^2$, where $L$ is the typical OC size and $D$ is the distance to the back gate. 
With this relation, the charging energy is of the order $E_C \sim e^2/C \sim e^2D/(\epsilon L^2) \lesssim 0.1 \text{meV} \approx1 \text{K}$, for a typical semiconducting dielectric constant $\epsilon \approx 10$. By contrast, the energy relaxation time, $\tau_E$, is proportional to the OC inverse level spacing, $\delta E_{}^{-1}$, when the OC is viewed as a 3D metallic island: $\tau_E =C_E[T_{\rm oc}(t)]/(\kappa_0 T_{\rm oc}) \sim  \delta E_{}^{-1}$, where $C_E[T_{\rm oc}(t)]$ is the OC heat capacity. 
See Appendix~\ref{sec:Appendix_A} for the derivation of this relation. We further estimate the level spacing of a typical micrometer-sized OC to  $\delta E_{}  \approx 0.1- 1\mu$K. 
To achieve a larger level spacing, say $\delta E_{} \sim 1 \text{K}$ for a 3D metallic island with Fermi wave length $\lambda_\mathrm{F}\sim 10 \text{\AA}$, demands an OC size of $L\sim 10 \text{nm}$. For a 2D OC, a similarly large level spacing is reached for $L\sim 100 \text{nm}$. The above estimates are consistent with the charging energy $E_C \approx 0.3$K and level spacing $\delta E_{} \approx 0.2 \mu $K, reported~\cite{Sivre2018Feb} for micrometer-sized OCs. This level spacing is thus much smaller than the typical measurement temperature in the $10$mK range. 

The estimates for $\tau_C$ and $\tau_E$, together with typically measured mesoscopic time scales of electron-electron thermalization, $\tau_{\rm e-e}$~\cite{Pothier1997Nov,Katine1998Jan,Ferrier2004Dec,Pierre2003Aug,Song2013Apr,Song2015Apr}, and electron-phonon thermalization, $\tau_{\rm e-ph}$, at cryogenic temperatures (see Table~\ref{tab:timescales}) motivate us in this work to assume the following hierarchy, or separation, of time scales
\begin{align}
\label{eq:hierachy_time}
\tau_C \ll \tau_{\rm e-e} \ll \tau_E \ll \tau_{\rm e-ph}.
\end{align}
Here, the first inequality, $\tau_C\ll \tau_{\rm e-e}$ states that the response of the OC's \textit{electrical potential} $V_{\rm oc}(t)$ is faster than the electron thermalization $\tau_{\rm e-e}$. 
This has the important consequence that the OC is in an out-of-equilibrium state on the time scale of the charge-response---a crucial ingredient to the heat Coulomb blockade effect of interest in this work.
Second, $\tau_{\rm e-e}\ll \tau_E$ states that the emitted energy fluctuations from the OC are given by a local and time-local, thermal state at temperature $T_{\rm oc}(t)$ at the time scale of energy fluctuations. This allows us to treat \emph{fluctuating} temperatures.
We further assume that the electron-phonon relaxation time $ \tau_{\rm e-ph}$ is the largest relevant time scale~\cite{Giazotto2006Mar}. This implies that electronic heat exchange with phonons can be neglected, which is in line with relevant experiments~\cite{Jezouin2013Nov,Srivastav2021May}. If the electron-phonon relaxation time is sizable, phonons enhance the system's heat conductance~\cite{Dashti2018Aug} (see also Eq.~\eqref{eq:energycons} below). 

As we discussed above, the separation $\tau_C\ll\tau_E$ is reasonable since a very short energy relaxation time $\tau_E$ becomes relevant only for an extremely small OC, and consequences of this can be detected only at very high measurement frequencies. However, in the model we present in this work, considering the opposite limit $\tau_E\ll\tau_C\rightarrow \infty$ may be relevant in the exotic but intriguing case where only fully charge-neutral particles (e.g., neutral modes in the fractional QH regime~\cite{Kane1994Jun,Bid2010Jul,Kumar2022Jan}) are injected into the OC. We postpone studies of this possibility to the future.  

Estimates for all experimentally relevant time scales, their corresponding energy scales, and their impact on the studied observables in this work are summarized in Table~\ref{tab:timescales}.

\subsection{Charge and heat dynamics - Langevin approach}
\label{sec:BL_approach}
The hierarchy~\eqref{eq:hierachy_time}, together with the fact that temperatures $T_{\rm oc},T_{\rm in} \gg \tau_E^{-1} $, motivates us to use a Langevin  approach to study the edge channel-OC system. Otherwise, for $T_{\rm oc}$ of similar magnitude as the OC level spacing, an accurate description demands instead a fully coherent quantum mechanical treatment, along the lines of Ref.~\cite{vandenBerg2015Jul}.  

Our Langevin analysis of the system in Fig.~\ref{fig:Single_Channel_Setup} starts by considering charge conservation, which establishes the following relation between the temporal OC charge $Q(t)$, and the incoming and outgoing edge charge currents $I_{\rm in}(t)$ and $I_{\rm out}(t)$, respectively:
\begin{align}
    \frac{dQ(t)}{dt} &= I_{\text{in}}(t) - I_{\text{out}}(t)\label{eq:Kirchhoff}.
\end{align}
Equation~\eqref{eq:Kirchhoff} is nothing but Kirchoff's current law describing the rate of change of the OC charge in terms of incoming and outgoing currents. The output charge current fluctuates with contributions from two sources: First, there are fluctuations coming from the fluctuating OC electrical potential $ \Delta V_{\rm oc}(t)\equiv \Delta Q(t)/ C$. Second, fluctuations are induced also due to the local OC temperature, which we model as a thermal equilibrium Langevin source, denoted $\delta I_{\rm oc}(t)$ (its correlations are given in Eq.~\eqref{eq:S_eq} below).  We thus have
\begin{align}
    \delta I_{\text{out}}(t) &= \frac{\Delta Q(t)}{\tau_C}  + \delta I_{\rm oc}(t).\label{eq:Langevin}
\end{align}
By definition, the fluctuations of any quantity, in particular here of the OC charge as well as the thermally induced current vanish on average: $\langle \delta I_{\rm oc}(t) \rangle =0 = \langle\Delta Q(t)\rangle$. 

Next, we note that for a single 1D channel with linear energy dispersion, which is an appropriate description at sufficiently low energy~\cite{Halperin1982Feb}, the heat current operators, $J$, and charge current operators, $I$, are generically related by the identity
\begin{align}
\label{eq:JI_relation}
    J = \frac{R_q}{2} I^2.
\end{align}
This identity is most easily derived with the bosonization technique (see Appendix~\ref{sec:Appendix_B}) and holds in standard fermionic scattering theory~\cite{Blanter2000Sep} when the energy-dependence of the velocity of scattering states is neglected. We would like to emphasize that Eq.~\eqref{eq:JI_relation} does not involve normal ordering (see, e.g., Ref.~\cite{vonDelft1998Nov}), which we instead take care of when we compute observables. Crucial for this work is that the identity~\eqref{eq:JI_relation} permits us to exactly compute the average heat current carried by the chiral channels, its charge-current noise and its heat-current noise, by using only charge-current correlation functions~\footnote{The relation~\eqref{eq:JI_relation} holds as long as the heat and charge is carried by the same type of particles, which is the case in this work.}.

However, to account for heat transport in the presence of OC temperature fluctuations, the relation~\eqref{eq:JI_relation} is not sufficient, and we also need to consider fluctuations in the OC energy $U(t)$. We therefore complement Eqs.~\eqref{eq:Kirchhoff}-\eqref{eq:JI_relation} with energy conservation
 \begin{align} \label{eq:energycons}
    \frac{dU(t)}{dt} = J^{\rm tot}_{\text{in}}(t) - J^{\rm tot}_{\text{out}}(t).
\end{align}
Here, $J^{\rm tot}_{\rm in}(t)$ and $J^{\rm tot}_{\rm out}(t)$ are the total, time-resolved incoming and outgoing heat currents in the edge channels. Note that a non-negligible phonon contribution would add to the right-hand side of Eq.~\eqref{eq:energycons}. The phonon contribution scales as $\sim T_{\rm oc}^p$ for $p>2$. For low temperatures, we assume this phonon contribution to be negligible (see experimental results underlining this assumption for temperatures $T_{\rm oc}\lesssim 70 \text{mK}$ in GaAs~\cite{Jezouin2013Nov} and for $T_{\rm oc}\lesssim 60 \text{mK}$ in graphene~\cite{Srivastav2021May}).

We now exploit the separation of time scales in Eq.~\eqref{eq:hierachy_time}. This permits us to write the following Langevin equation for the total output heat-current fluctuations
\begin{align}
\label{eq:heatfluct}
    \delta J^{\rm tot}_{\text{out}}(t) = \frac{\Delta U(t)}{\tau_E} + \delta J_{\text{out}}^{}(t),
\end{align} 
which thus have two distinct contributions. The fluctuations in the internal energy, in the presence of OC temperature fluctuations and a finite OC heat capacity, emerge on the time scale $\tau_E$ that is much longer than the scale governing the charge fluctuations, see Eq.~\eqref{eq:hierachy_time}. The energy fluctuations can be written within the Langevin approach as 
\begin{align}\label{eq:energyconsfluc}
    \Delta U(t) \approx C_E[T_{\rm oc}]  \Delta T_{\rm oc}(t),
\end{align}
where we kept $\Delta T_{\rm oc}$ only to linear order~\footnote{From Eq.~\eqref{eq:temp_supp} we see that the approximation of linear order temperature fluctuations holds well when the typical OC temperature fluctuations are small, i.e., for $\sqrt{\left\langle \left\langle  \Delta T_{\rm oc}(0)^2 \right\rangle \right\rangle_E}/T_{\rm oc}\ll 1$. Inserting the estimated OC level spacing from Appendix~\ref{sec:Appendix_A} gives  
$\delta E \ll \frac{\pi^2}{2}  T_{\rm oc}$ which for a level spacing $  \delta E \approx 0.1 \mu K - 1 \mu K$ is well exceeded by typical experimental temperatures $T_{\rm oc} \approx 8$mK, see, e.g., Ref.~\cite{Sivre2018Feb}.} and thus take the heat capacity $C_E[T_{\rm oc}(t)]\approx C_E[T_{\rm oc}]$, i.e., it depends only on the average OC temperature $T_{\rm oc}$. The impact of higher orders of $\Delta T_{\rm oc}(t)$ are negligible due to the large $\tau_E$. Combining Eq.~\eqref{eq:heatfluct} and Eq.~\eqref{eq:energyconsfluc}, we obtain a linearized Langevin equation for the total output heat-current fluctuations as
\begin{align}
\label{eq:linearized_total_heat_fluct}
    \delta J^{\rm tot}_{\text{out}}(t) = \kappa T_{\rm oc} \Delta T_{\rm oc}(t) + \delta J^{}_{\text{out}}(t).
\end{align}
Here, $\kappa T_{\rm oc}\equiv C_E[T_{\rm oc}]/(\tau_E T_{\rm oc})\times T_{\rm oc}$ is the output heat conductance (to be computed below). The heat-current fluctuations $\delta J_{\text{out}}^{}(t)$ are directly related via Eq.~\eqref{eq:JI_relation} to the total charge-current fluctuations in Eq.~\eqref{eq:Langevin} and take place on the time scale $\tau_C$, which is much shorter than the energy relaxation time $\tau_E$.

The heat-current fluctuations in Eq.~\eqref{eq:heatfluct} and \eqref{eq:linearized_total_heat_fluct}, are hence the fluctuations with respect to a heat current obtained from a double averaging procedure, $\delta J^{\rm tot}_{\text{out}}(t) \equiv  J^{\rm tot}_{\text{out}}(t) - \left\langle \left\langle J^{\rm tot}_{\text{out}}(t)\right\rangle \right\rangle_E $, namely by averaging with respect to the (fast) charge fluctuations, denoted by simple brackets as $\left\langle \cdots \right\rangle$, and with respect to the (slow) temperature fluctuations, denoted as $\left\langle \cdots \right\rangle_E$. With this averaging procedure, the average rate of change of $U$ equals the heating/cooling power applied to the OC, e.g., Joule heating or cooling by phonons. 

\section{Single edge channel dynamics}
\label{sec:SEC}
\subsection{Charge current, charge-current noise, heat current, and effective output temperature}
\label{sec:Single_channel_charge_sector}
For completeness, in this subsection we follow Ref.~\cite{slobodeniuk_equilibration_2013} closely, and recapture the steps to determine the output charge current and charge-current noise in the standardly considered limit where $\tau_E\to\infty$. 
We discuss how one can determine the average heat current of the generically out-of-equilibrium output channel, define a corresponding effective output temperature, and show how it is affected by the heat Coulomb blockade.

\subsubsection{Output charge current and charge-current noise}
To solve the output charge-current dynamics, Eqs.~\eqref{eq:Kirchhoff} and~\eqref{eq:Langevin}, it is convenient to do the analysis in frequency space, $I_{\rm out}(\omega)$. The solution takes the form of a linear combination of the input and OC currents:
\begin{align}
    &I_{\rm out}(\omega) = \sum_{p=\rm{in},\rm{oc}}\mathcal{T}_p(\omega) I_{p}(\omega),\label{eq:jSol}\\
    & \mathcal{T}_{\rm in}(\omega) = 1-\mathcal{T}_{\rm oc}(\omega) = \left[1-i\omega  \tau_C\right]^{-1}\label{eq:t_values}.
\end{align}
Here, the frequency-dependent coefficients $\mathcal{T}_{p}(\omega)$, with $p\in \{\text{in},\text{oc}\}$, depend also on the $RC$ time $\tau_C$, and can be viewed as scattering amplitudes of bosonic density fluctuations~\cite{Safi1999Dec} carried along the chiral edge channel and by excitations inside the OC, see Appendix~\ref{sec:Appendix_B} for details. Charge current conservation together with Eqs.~\eqref{eq:jSol} and~\eqref{eq:t_values} leads to equal average input and output charge currents
\begin{align}
\label{eq:I_out_I_in}
    I_{\rm out}(\omega=0) = I_{\rm in}(\omega=0).
\end{align}
The finite-frequency output charge-current noise $S_{\rm out}(\omega)$---here, presenting its non-symmetrized form for compactness---is defined as the two-point correlation function of the output fluctuations $\delta I_{\rm out}(\omega)\equiv I_{\rm out}(\omega)-\langle I_{\rm out}(\omega) \rangle$,
\begin{align}
\label{eq:S_def}
    \langle \delta I_{\rm out}(\omega) \delta I_{\rm out}(\omega') \rangle \equiv 2\pi \delta(\omega+\omega')S_{\rm out}(\omega).
\end{align}
The OC Langevin source, $\delta I_{\rm oc}(t)$, results from the thermally induced uncertainty in the OC occupation and is hence described by a \textit{local equilibrium}. Therefore, all averages $\langle ... \rangle$ are taken with respect to the product of the thermal input state (note that the input channel emanates from a large Fermi reservoir in equilibrium) and the thermal OC Langevin source state. By combining the output-current solution~\eqref{eq:jSol} and the definition for the charge-current noise~\eqref{eq:S_def}, one finds the out-of-equilibrium finite-frequency charge-current noise 
\begin{subequations}
 \label{eq:Sout}
\begin{align}
    S_{\rm out}(\omega)& =  \sum_{p=\rm{in},\rm{oc}}|\mathcal{T}_p(\omega)|^2 S_p(\omega)\\
   & = \frac{S_{\rm in}(\omega) +S_{\rm oc}(\omega)\omega^2 \tau_C^2}{1+\omega^2 \tau_C^2}\\
    &\to\left\{
    \begin{array}{ll}
       S_\mathrm{in}(\omega)   & \omega\tau_C\ll1, \\
        S_\mathrm{oc}(\omega)   & \omega\tau_C\gg1, 
    \end{array}
    \right.
\end{align}    
\end{subequations}
where $S_p(\omega)$ is the charge-current noise of subsystem $p$, given in the equilibrium form as
\begin{subequations}
\label{eq:S_eq}
\begin{align}
\langle \delta I_p(\omega) \delta I_p(\omega') \rangle & \equiv 2\pi \delta(\omega+\omega')S_p(\omega),\\
    S_p(\omega) &=  \frac{ \omega R_q^{-1}}{1-e^{- \omega/T_p}}.
\end{align}
\end{subequations}
Here, $T_p$ is the input channel temperature for $p=\rm{in}$ and it is the OC temperature for $p=\rm{oc}$. It follows from the scattering amplitude definitions~\eqref{eq:t_values} that
\begin{align}
\label{eq:SumRule}
\sum_{p=\rm{in},\rm{oc}}|\mathcal{T}_p(\omega)|^2 = 1,
\end{align}
which reflects charge conservation. It is important to point out that $S_{\rm out}(\omega)$ in Eq.~\eqref{eq:Sout} is generally not on equilibrium form~\eqref{eq:S_eq} but will be so if either $T_{\rm in}=T_{\rm oc}$ or in the limits $\omega \tau_C \rightarrow 0,\infty$, to be discussed further in Sec.~\ref{eq:temp_comparisons} below.  Equation~\eqref{eq:Sout} thus indicates that the OC acts as a band-pass filter for the chiral edge channel, i.e., the value of the parameter $\omega \tau_C$ determines whether it is the input channel or the OC that dominates in the output noise contribution.

\subsubsection{Effective output temperature and output heat current}
The output charge-current noise~\eqref{eq:Sout} can be used to establish an effective temperature of the output channel, which is generally not in thermal equilibrium. 
One possibility would be to parallel the Johnson-Nyquist relation~\cite{Nyquist1928,Johnsson1928} and define a Johnson-Nyquist temperature $T^{\rm JN}_{\rm out}\equiv R_q S_{\rm out}(0)$ with Eq.~\eqref{eq:Sout}. 
Here, taking the zero-frequency limit amounts to considering frequencies $\omega \tau_C,\omega/T_p \ll 1$, leading to
\begin{align}
\label{eq:ZeroFreqTemp}
    T^{\rm JN}_{\rm out}&= R_q S_{\rm out}(0) = R_q \sum_{p=\rm{in},\rm{oc}}\mathcal{T}_p(0) S_p(0) \notag \\
    &= R_q (R_q^{-1}T_{\rm in}+0) = T_{\rm in}.
\end{align}
Physically, this result is a consequence of the OC being floating and that only a single, chiral edge channel is connected. Charge conservation then demands that the low-frequency output fluctuations must equal the low-frequency input fluctuations [see Eq.~\eqref{eq:I_out_I_in}] which are indeed characterized by $T_{\rm in}$.  In other words, charge-current fluctuations averaged over very long time scales do not ``see'' effects of the OC charge capacitance $C$. As an effective temperature taking account OC effects, Eq.~\eqref{eq:ZeroFreqTemp} is hence not a very useful quantity. 
Note, however, that when taking the limit $\omega/T_p\ll1$, but $\omega \tau_C\gg1$ in Eq.~\eqref{eq:Sout}, one finds instead $T^{\rm JN}_{\rm out}=R_qS_{\rm oc}(0)=T_{\rm oc}$~\footnote{Alternatively, the behaviour of the low frequency charge noise can be understood by comparing the observation (or measurement) time, $\tau_{\rm obs}$ with $\tau_C$. Explicitly, we have from Eq.~\eqref{eq:Sout} that
$S_{\rm out}(0) \approx \int_{-\tau_{\rm obs}/2}^{\tau_{\rm obs}/2} dt  S(t)$ $=\int_{-\infty}^\infty dx  \frac{\sin(x)}{\pi x} \frac{S_{\rm in}(\frac{2 x}{\tau_{\rm obs} }) +S_{\rm oc}(\frac{2x}{\tau_{\rm obs} })\frac{4 x^2 \tau_C^2}{\tau_{\rm obs}^2}}{1+\frac{4 x^2 \tau_C^2}{\tau_{\rm obs}^2}}$, where we assumed $\tau_{\rm obs} T_p \gg 1$, so that we can set $S_{p}(\omega)\approx S_{p}(0)$. For $\tau_{\rm obs} \gg \tau_C$, we recover Eq.~\eqref{eq:ZeroFreqTemp}, while in the opposite limit $\tau_{\rm obs} \ll \tau_C$, we find instead $T^{\rm JN}_{\rm out}=R_qS_{\rm oc}(0)=T_{\rm oc}$ as expected.}. Physically, this special limit means that for an infinite OC charge capacitance, any information carried by the input edge state, such as its injection temperature, is lost upon particles entering and then exiting the OC. The OC then acts like a proper thermal reservoir for the output channel. 

The shortcoming of the Johnson-Nyquist temperature~\eqref{eq:ZeroFreqTemp} demands an alternative 
definition of the effective output temperature, that captures features of the noise over a broader frequency range. This can be achieved by considering the following equilibrium relation between the 1D heat current~\cite{Pendry1983Jul} $J_{\rm eq}$ and the temperature $T$,
\begin{align}
\label{eq:JT_rel}
    J_{\rm eq} &=  \int_0^\infty \frac{d\omega}{2\pi} \omega n_B(\omega) =\frac{R_q}{4\pi}\int_{-\infty}^\infty d\omega \left[ S(\omega)-\frac{\omega\theta(\omega)}{R_q}\right] \notag \\
    &=\frac{\pi T^2}{12} \equiv \frac{\kappa_0}{2}T^2.
\end{align} 
Here, the term $\omega\theta(\omega)/R_q$, where $\theta(\omega)$ is a step function, can be interpreted as a subtraction of vacuum fluctuations (the $T=0$ contribution),
\begin{align}
\label{eq:S_zero_T}
    S_p(\omega) = \frac{R_q^{-1}\omega}{1-e^{-\omega/T_p}} \rightarrow \omega R_q^{-1} \theta(\omega).
\end{align} 
The edge channel heat current is thereby expressed via the thermal energy carried by the bosonic fluctuations, given in terms of the Bose-Einstein distribution
\begin{equation}
\label{eq:NB_def}
    n_B(\omega)\equiv \frac{1}{\exp(\omega/T)-1}.
\end{equation}
In the second equality in Eq.~\eqref{eq:JT_rel}, one further uses that $n_B(\omega)$ is related to the equilibrium noise~\eqref{eq:S_eq} as
\begin{align}
\label{eq:nb_S_relation}
    \omega \left[1+n_{B}(\omega)\right] =R_q S(\omega).
\end{align}
 The relation~\eqref{eq:JT_rel} holds in equilibrium for any abelian 1D chiral channel, charged or not, fractional or not, since the heat current does not depend on $R_q$.

One can now \textit{define} an effective temperature as the analogous expression of~\eqref{eq:JT_rel} but for a \textit{generic charge-current noise} (i.e., not requiring equilibrium). For the setup in Fig.~\ref{fig:Single_Channel_Setup}, this yields the output heat current as the integral over the output charge-current noise 
\begin{align}
\label{eq:T_integral}
    J_{\rm out} = \frac{R_q}{4\pi}\int_{-\infty}^{\infty}d\omega \left[S_{\rm out}(\omega )-\frac{\omega\theta(\omega)}{R_q}\right] \equiv \frac{\kappa_0}{2}T_{\rm out}^2.
\end{align}
For the output charge-current noise~\eqref{eq:Sout}, the sum rule~\eqref{eq:SumRule} ensures convergence of the integral in Eq.~\eqref{eq:T_integral}. Moreover, Eq.~\eqref{eq:T_integral} is consistent with the heat current obtained by integrating the, generally out-of-equilibrium, output \textit{electronic} distribution, see Appendix~\ref{sec:Appendix_equivalence}.

To explicitly compute $T_{\rm out}$ for the system in Fig.~\ref{fig:Single_Channel_Setup}, one thus inserts the output noise~\eqref{eq:Sout} into the integral~\eqref{eq:T_integral}, and substitutes the scattering amplitudes~\eqref{eq:t_values} and the equilibrium noises~\eqref{eq:S_eq}. After a series of algebraic manipulations, this yields
\begin{align}
\label{eq:T-evaluated}
    T^2_{\rm out} = T^2_{\rm oc}+\frac{6}{\tau_C^2\pi^2}\Big[ \mathcal{F}\left( \frac{1}{\tau_C T_{\rm in}}\right)-\mathcal{F}\left( \frac{ 1}{\tau_C T_{\rm oc}}\right)\Big],
\end{align}
which is expressed in terms of the dimensionless integral function
\begin{align}
\label{eq:integral_function}
    \mathcal{F}(a) &\equiv \int_0^\infty \frac{z}{z^2+a^2}\frac{1}{e^{z}-1}dz \notag\\  &= \frac{1}{2}\Big[ \ln \left( \frac{a}{2\pi}\right)-\frac{\pi}{a}-\psi\left( \frac{a}{2\pi}\right)  \Big].
\end{align}
Here, $\psi(z)$ is the logarithmic derivative of the gamma-function. Two useful limits of $\mathcal{F}(a)$ are
\begin{subequations}
\label{eq:F_limits}
\begin{align}
    &\mathcal{F}(a\gg 1)\approx \frac{\pi^2}{6a^2} - \frac{\pi^4}{15 a^4},\\
    &\mathcal{F}(a\ll 1)\approx \frac{\pi}{2a}.
\end{align}
\end{subequations}
Equation~\eqref{eq:T-evaluated} demonstrates that $T_{\rm out}$ depends explicitly on the competition between the $RC$ time $\tau_C$ and the thermal time scales set by the input and OC temperatures $T_{\rm in}$ and $T_{\rm oc}$.

\subsubsection{Single-channel heat Coulomb blockade}
\label{sec:SCHC}
The implication of Eq.~\eqref{eq:T-evaluated} is most conveniently analyzed in two important limits: (i) a very cold OC, $T_{\rm oc}\ll T_{\rm in}$, and (ii) a very cold input channel $T_{\rm in}\ll T_{\rm oc}$. Taking these two limits amounts to setting $\tau_C T_p\rightarrow 0$, for $p=\rm{oc}$ and $p=\rm{in}$, respectively, in Eq.~\eqref{eq:T-evaluated}.
For configuration (i), by using the asymptotics~\eqref{eq:F_limits}, we thus find 
\begin{align}
\label{eq:Cold_Contact_limits}
    \frac{T_{\rm out}}{T_{\rm in}} \approx \begin{cases}
  \sqrt{1-3\pi^2 (\tau_C T_{\rm in})^2/5}  & \tau_C T_{\rm in}\ll  1, \\
  \sqrt{3/(\pi\tau_C T_{\rm in})} & \tau_C T_{\rm in}\gg  1.
\end{cases}
\end{align}
Here, the small correction to $1$ for $\tau_C T_{\rm in}\ll 1$ has not been previously reported, and thus extends the result in Ref.~\cite{slobodeniuk_equilibration_2013}. 
From Eq.~\eqref{eq:Cold_Contact_limits}, we see that for sufficiently small $\tau_C T_{\rm in}$, the effective temperature of the output signal approaches the \textit{input} temperature, $T_{\rm out}=T_{\rm in}$. It hence seems that the electron distribution of the input channel does hardly interact with the OC. This shows how the OC acts as a bandpass filter, resulting in a type of ``teleportation" effect~\cite{Idrisov2018Jul,Duprez2019}, hindering a (full) equilibration of the input signal with the OC distribution.
 By contrast, for large $\tau_C T_{\rm in}$, the input and output temperatures become related as $T_{\rm out} \propto \sqrt{T_{\rm in}}$. The limit $T_\mathrm{out}=T_\mathrm{oc}=0$ is then furthermore reached when $\tau_C T_\mathrm{in}\to\infty$ for fixed input temperature $T_\mathrm{in}$, namely when the OC charge response becomes very slow and the equilibration becomes efficient.
\begin{figure}[t]
\begin{center}
\captionsetup[subfigure]{position=top,justification=raggedright}
\subfloat[]{
\includegraphics[width=0.95\columnwidth]{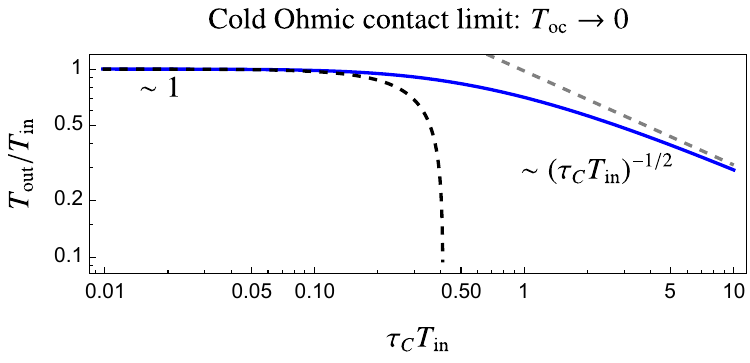}
\label{fig:Chargenoise_cold_contact}}
\\[-0.25cm]
\subfloat[]{
\includegraphics[width =0.95\columnwidth]{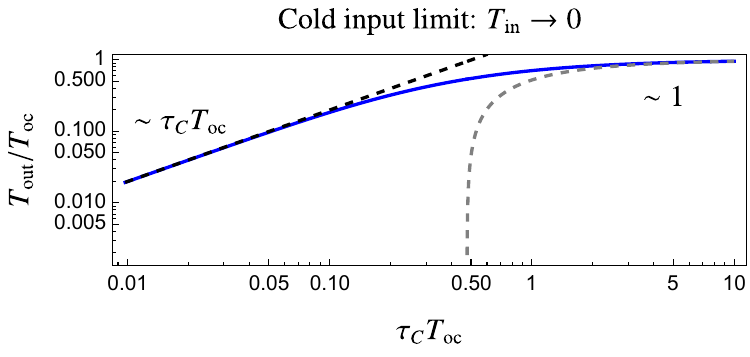}
\label{fig:Chargenoise_cold_input}}
\caption{(a) Blue, solid line: Ratio of output and input effective temperatures, $T_{\rm out}/T_{\rm in}$, vs $\tau_C T_{\rm in}$.  The dashed, gray and black lines are the analytically obtained limits~\eqref{eq:Cold_Contact_limits}. The lines are shown in the limit of a cold OC, $\tau_CT_{\rm oc}\ll 1$. (b)~Blue, solid line: Ratio of output and OC effective temperatures, $T_{\rm out}/T_{\rm oc}$, vs $\tau_C T_{\rm oc}$. The dashed, gray and black lines are the analytically obtained limits~\eqref{eq:Cold_Input_limits}. The lines are shown in the limit of a cold input channel, $\tau_CT_{\rm in}\ll 1$.}
\label{fig:Chargenoises}
\end{center}
\end{figure}

For configuration (ii), by using the asymptotic limits~\eqref{eq:F_limits} for the output temperature of Eq.~\eqref{eq:T-evaluated}, we instead find that 
\begin{align}
\label{eq:Cold_Input_limits}
    \frac{T_{\rm out}}{T_{\rm oc}} \approx \begin{cases}
  \sqrt{\frac{2}{5}}\pi\tau_C T_{\rm oc}  & \tau_C T_{\rm oc}\ll 1, \\
  \sqrt{1-3/(\pi \tau_C T_{\rm oc})} & \tau_C  T_{\rm oc}\gg  1.
\end{cases}
\end{align}
Also here, the asymptotic limit for $\tau_C T_{\rm oc}\ll 1$ has not been reported before and it extends the results in Ref.~\cite{slobodeniuk_equilibration_2013}. Again, the ability of the OC to equilibrate the cold input state is impeded (i.e., one finds $T_{\rm out} \rightarrow T_\mathrm{in}\to0$) due to the bandpass filter or teleportation effect, in this case for sufficiently small $\tau_C T_{\rm oc}$. For large $\tau_C T_{\rm oc}$, one finds $T_{\rm out}\approx T_{\rm oc}$ so that the equilibration instead is very efficient. 

We plot the limits~\eqref{eq:Cold_Contact_limits} and~\eqref{eq:Cold_Input_limits} together with the full expression~\eqref{eq:T-evaluated} in Figs.~\ref{fig:Chargenoise_cold_contact} and~\ref{fig:Chargenoise_cold_input}, respectively. The impeded equilibration effect that is displayed here lies at the heart of the heat Coulomb blockade effect~\cite{Duprez2019}, see also Sec.~\ref{sec:Multi_Gen_Charge} for the multi-channel case, below. 

\subsection{Impact of potential fluctuations on the heat-current noise}
\label{eq:HeatNoise_Voltage}

We now go beyond the analysis in Sec.~\ref{sec:Single_channel_charge_sector}, most of which was previously reported in Ref.~\cite{slobodeniuk_equilibration_2013}, and investigate the heat current fluctuations of the output channel and how they are impacted by the heat Coulomb blockade effect. We first consider the commonly studied regime, where $\tau_E\to\infty$ and $T_\mathrm{oc}$ is hence fixed. 

\subsubsection{Relation between finite-frequency charge- and heat-current noise}
Our starting point here is the observation that, for charge-current noise, $S(\omega)$, and heat-current noise, $S^J(\omega)$, defined as
\begin{subequations}
\begin{align}
    & \left\langle \delta I_{\rm }(\omega) \delta I^{}_{\rm }(\omega') \right\rangle  \equiv 2\pi \delta(\omega+\omega') S_{\rm }(\omega),\\
    & \left\langle \delta J_{\rm }(\omega) \delta J^{}_{\rm }(\omega') \right\rangle  \equiv 2\pi \delta(\omega+\omega') S^J_{\rm }(\omega),
\end{align}
\end{subequations}
the charge-heat relation~\eqref{eq:JI_relation} implies the following relation (see Ref.~\cite{Idrisov2022Sep} for a derivation)
\begin{align}
\label{eq:SQ}
    S^J(\omega) = \frac{R_q^2}{4\pi}\int_{-\infty}^{\infty} d\omega_1 S(\omega_1)S(\omega-\omega_1).
\end{align}
Performing the integral for $S(\omega)$ on the equilibrium form~\eqref{eq:S_eq} gives 
\begin{align}
\label{eq:Sp_gen}
    S^J(\omega)  = \frac{\omega}{48\pi}\left[(2\pi T)^2+\omega^2\right]\left(1+\coth\left(\frac{\omega}{2T}\right)\right),
\end{align}
in agreement with the equilibrium heat-current noise derived in Ref.~\cite{Averin2010Jun}. The two important limits of $S^J(\omega)$ are for $\omega/T\ll 1$ and $\omega /T \gg 1$, producing $S^J(\omega)\sim T^3$ and $S^J(\omega)\sim |\omega|^3$, respectively. 

The relation~\eqref{eq:SQ} is valid for a chiral 1D edge channel with linear dispersion. Since Eq.~\eqref{eq:JI_relation} is an operator identity, this identity and, in turn, Eq.~\eqref{eq:SQ} hold for any edge channel---and in particular for the output channel. We thus obtain the output heat-current noise by inserting into~\eqref{eq:SQ} the  output charge-current noise $S_{\rm out}(\omega)$ from Eq.~\eqref{eq:Sout} and find
\begin{align}
\label{eq:SQout}
     S_{\rm out}^{J}(\omega) &= \frac{R_q^2}{4\pi}\int_{-\infty}^{\infty} d\omega_1 \sum_{p=\rm{in},\rm{oc}}|\mathcal{T}_p(\omega_1)|^2 S_p(\omega_1) \notag \\ & \times \sum_{k=\rm{in},\rm{oc}}|\mathcal{T}_k(\omega-\omega_1)|^2 S_k(\omega-\omega_1).
\end{align}
For the remainder of this subsection, we focus on the low-frequency heat-current noise $S_{\rm out}^{J}(0)$, i.e., frequencies $\omega\ll T_{\rm in},T_{\rm oc},\tau_C^{-1}$. As a first check, we consider the global equilibrium case for which $T_{\rm in}=T_{\rm oc}=T$. Using these parameters in the equilibrium charge-current noise~\eqref{eq:S_eq} and the sum rule~\eqref{eq:SumRule}, Eq.~\eqref{eq:SQout} reduces to
\begin{align} \label{eq:SQ_equilibrium}
     S_{\rm out}^{J}(0) &= \frac{1}{4\pi}\int_{-\infty}^{\infty} d\omega_1 \frac{\omega_1}{1-e^{-\omega_1/T}} \frac{-\omega_1}{1-e^{\omega_1/T}} =\kappa_0T^3,
\end{align}
which is indeed the equilibrium heat-current noise~\cite{Averin2010Jun}. The physical interpretation of the absence of $\tau_{C}$ in Eq.~\eqref{eq:SQ_equilibrium} is simply that for a global, uniform temperature $T$, the $RC$ time of the OC is completely immaterial for the heat-current noise, since the particle distributions are always at the same temperature, regardless of the OC charge relaxation time.

\subsubsection{An alternative effective temperature}
At this point, we note that the relation~\eqref{eq:SQ_equilibrium} suggests an alternative definition for an effective output temperature, 
\begin{align}
\label{eq:T_def_new}
    \tilde{T}^3_{\rm out} \equiv \frac{S^J_{\rm out}(0)}{\kappa_0},
\end{align}
which parallels the definition~\eqref{eq:T_integral}. While it is clear that the temperature definitions~\eqref{eq:T-evaluated}~and~\eqref{eq:T_def_new} agree in equilibrium, so that $\tilde{T}_{\rm in}=T_{\rm in}$ and $\tilde{T}_{\rm oc}=T_{\rm oc}$, we now show that this is not the case for the \textit{out-of-equilibrium} output channel.
\begin{figure}[t]
\begin{center}
\captionsetup[subfigure]{position=top,justification=raggedright}
\subfloat[]{
\includegraphics[width=0.95\columnwidth]{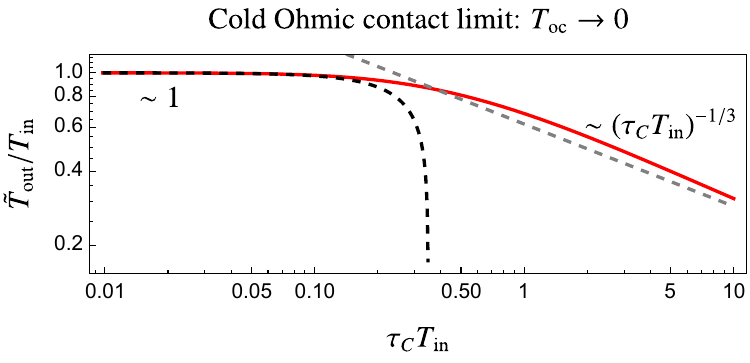}
\label{fig:Heatnoise_cold_contact}}
\\[-0.25cm]
\subfloat[]{
\includegraphics[width =0.95\columnwidth]{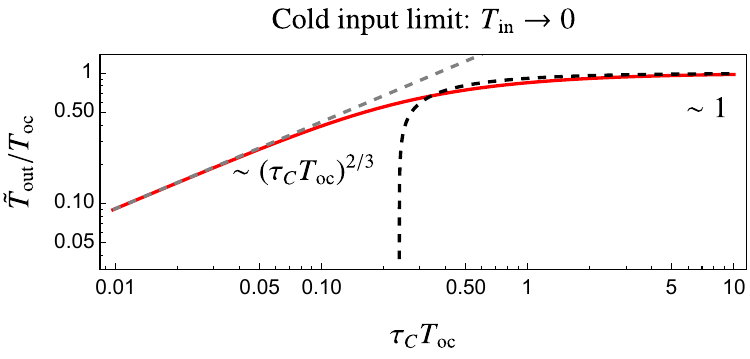}
\label{fig:Heatnoise_cold_input}}
\caption{(a) Red, solid line: Ratio of output and input temperatures vs $\tau_C T_{\rm in}$, where $\tau_C=R_q C$ is the $RC$ time and $T_{\rm in}$ is the input temperature. The plots are made for the limit of a cold OC $\tau_CT_{\rm oc}\ll 1$. The dashed, gray and black lines are the analytically obtained limits~\eqref{eq:SQout_asymptotic_1}. (b) Red, solid line: Ratio of output and OC temperatures vs $\tau_C T_{\rm oc}$, where $T_{\rm oc}$ is the OC temperature. The plots are made for the limit of a cold input channel $\tau_C T_{\rm in}\ll 1$. The dashed, gray and black lines are the analytically obtained limits~\eqref{eq:SQout_asymptotic_2}.}
\label{fig:Heatnoises}
\end{center}
\end{figure}

To this end, we consider the same two configurations as in the previous subsection. For configuration (i), we take in Eq.~\eqref{eq:T_def_new} the limit $T_{\rm oc}\ll T_{\rm in}$ and obtain the integral expression
\begin{align}
\label{eq:SQout_integral1}
    S_{\rm out}^J(0) & = \notag \\ & \frac{T_{\rm in}^3}{2\pi} \int_{0}^{\infty }dz \frac{e^z z^2 \left(\left(\tau_C T_{\rm in}\right)^2 z^2+1\right)-\left(\tau_C T_{\rm in}\right)^2 z^4}{\left(e^z-1\right)^2 \left(\left(\tau_C T_{\rm in}\right)^2 z^2+1\right)^2}.
\end{align}
In the two interesting limits $\tau_C T_{\rm in} \ll 1$ and $\tau_C T_{\rm in}\gg 1$, we find that Eq.~\eqref{eq:SQout_integral1} can be analytically approximated as
\begin{align}
\label{eq:SQout_asymptotic_1}
    \frac{S_{\rm out}^J(0)}{S_{\rm in}^J(0)} & = \frac{\tilde{T}^3_{\rm out}}{T^3_{\rm in}} \notag\\ &\approx \begin{cases}
   1 +\left(\frac{72  \zeta (5)}{\pi ^2}  -\frac{8 \pi ^2 }{5}\right) (\tau_C T_{\rm in})^2 &    \tau_C  T_{\rm in} \ll 1,\\
\frac{3}{4\pi} \left(\tau_C T_{\rm in}\right)^{-1} & \tau_C T_{\rm in} \gg 1,
\end{cases}
\end{align}
where $\zeta(z)$ is the Riemann zeta-function (with $\zeta(5)\approx 1.04$) and $S^J_{\rm in}(0)=\pi T^3_{\rm in}/6$ is the zero-frequency input heat-current noise. From Eq.~\eqref{eq:SQout_asymptotic_1}, we see that also the heat-current noise reveals that for $\tau_C T_{\rm in}\ll 1$, the OC cannot cool the input channel. For large $\tau_C T_{\rm in}$, we have $\tilde{T}_{\rm out}\sim T_{\rm in}^{2/3}$, which differs from $T_{\rm out}\sim \sqrt{T_{\rm in}}$ as obtained in Eq.~\eqref{eq:Cold_Contact_limits}. We plot the full integral~\eqref{eq:SQout_integral1} and the asymptotics~\eqref{eq:SQout_asymptotic_1} in Fig.~\ref{fig:Heatnoise_cold_contact}.

For configuration (ii), we take in Eq.~\eqref{eq:T_def_new} instead $T_{\rm oc}\gg T_{\rm in}$ and find
\begin{align}
\label{eq:SQout_integral2}
    S_{\rm out}^J(0) &= \notag \\ &\frac{T_{\rm oc}^3}{2\pi} \int_{0}^{\infty }dz \frac{\left(\tau_C T_{\rm oc}\right)^2 z^4 \left(e^z \left(\left(\tau_C T_{\rm oc}\right)^2 z^2+1\right)-1\right)}{\left(e^z-1\right)^2 \left(\left(\tau_C T_{\rm oc}\right)^2 z^2+1\right)^2}.
\end{align}
In the two limits $\tau_C T_{\rm oc} \ll 1$ and $\tau_C T_{\rm oc} \gg 1$ , this integral can be analytically approximated as
\begin{align}
\label{eq:SQout_asymptotic_2}
    \frac{S_{\rm out}^J(0)}{S_{\rm oc}^J(0)} = \frac{\tilde{T}^3_{\rm out}}{T^3_{\rm oc}} \approx \begin{cases}
  \frac{72\zeta(5)}{\pi^2}  \left(\tau_C T_{\rm oc} \right)^2& \tau_C T_{\rm oc}\ll 1, \\ 
  1 - \frac{3}{4\pi} \left(\tau_C T_{\rm oc}\right)^{-1} &   \tau_C T_{\rm oc} \gg 1,
  \end{cases}
\end{align}
where $S_{\rm oc}^J(0) = \pi T_{\rm oc}^3/6$ is the zero-frequency OC heat-current noise. We thus see that for $ \tau_C T_{\rm oc} \ll 1$, the OC cannot equilibrate the input channel. For small $\tau_C T_{\rm oc}$, we have $\tilde{T}_{\rm out}\sim T_{\rm oc}^{5/3}$, to be contrasted with $T_{\rm out}\sim T_{\rm oc}^2$ in Eq.~\eqref{eq:Cold_Input_limits}. 
We plot the numerically evaluated  integral~\eqref{eq:SQout_integral2} together with the asymptotics~\eqref{eq:SQout_asymptotic_2} in Fig.~\ref{fig:Heatnoise_cold_input}.

\begin{figure}[t!]
\begin{center}
\captionsetup[subfigure]{position=top,justification=raggedright}
\subfloat[]{
\includegraphics[width=0.95\columnwidth]{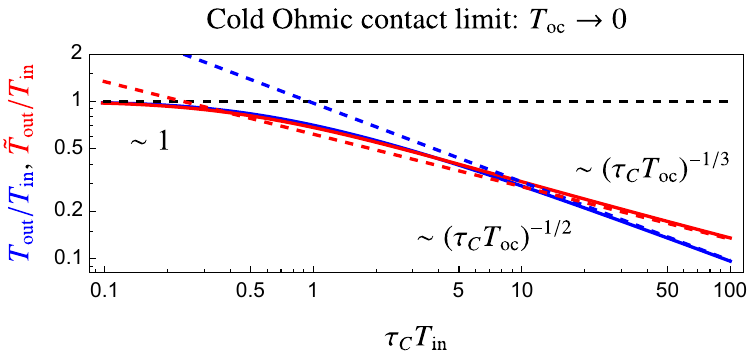}
\label{fig:temp_comparisons_cold_contact}}
\\[-0.25cm]
\subfloat[]{
\includegraphics[width =0.95\columnwidth]{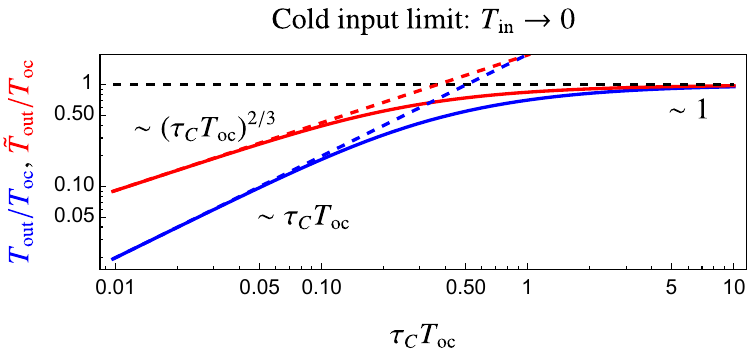}
\label{fig:temp_comparisons_cold_input}}
\caption{Comparison between the effective output temperatures $T_{\rm out}$ (solid blue lines) in Eq.~\eqref{eq:T-evaluated} and $\tilde{T}_{\rm out}$ (solid red lines) in Eq.~\eqref{eq:T_def_new}. (a) The temperatures are compared in the cold Ohmic contact limit, where $\tau_C T_{\rm oc}\ll 1$ and $T_{\rm in}$ is kept finite. The blue and red dashed lines are the asymptotic expressions~\eqref{eq:Cold_Contact_limits} and~\eqref{eq:SQout_asymptotic_1}, respectively, both for $\tau_C T_{\rm in}\gg 1$.  (b) Comparison in the cold input limit, where $\tau_C T_{\rm in}\ll 1$ and $T_{\rm oc}$ is finite. The blue and red dashed lines are the asymptotic expressions~\eqref{eq:Cold_Input_limits} and~\eqref{eq:SQout_asymptotic_2}, respectively, both for $\tau_C T_{\rm oc}\ll 1$. }
\label{fig:temp_comparisons}
\end{center}
\end{figure}

We have thus in this subsection demonstrated that heat Coulomb blockade (in the sense that the OC's finite charge capacitance impedes equilibration between the OC and the edge channel) is manifest also in the zero-frequency heat-current noise $S^{J}_{\rm out}(0)$. 
In turn, this quantity can be interpreted as an output temperature according to Eq.~\eqref{eq:T_def_new}. 
That not only the output charge-current noise but also the output heat-current noise is affected by $\tau_C$ is a direct consequence of the charge-heat relations~\eqref{eq:JI_relation} and~\eqref{eq:SQ}, valid for 1D chiral transport at low energies. The results of this subsection hold because we take into account the OC potential fluctuations exactly. 
The impact of $\tau_C$ on the heat-current noise is indeed absent to linear order in the OC potential fluctuations, as reported in Ref.~\cite{Dashti2018Aug}. 

Before we move to describing the full dynamics of the outgoing heat-current fluctuations, we compare the alternative effective-temperature definitions and we also demonstrate two additional, important and experimentally-relevant instances where the heat transport is affected by the OC potential fluctuations.

\subsubsection{Comparison of effective output temperatures}
\label{eq:temp_comparisons}
To compare the two output-temperature definitions $\tilde{T}_{\rm out}$ and $T_{\rm out}$ (Eq.~\eqref{eq:T-evaluated}~and~Eq.~\eqref{eq:T_def_new}, respectively), we plot in Fig.~\ref{fig:temp_comparisons} the full integral expressions~\eqref{eq:T-evaluated},~\eqref{eq:SQout_integral1}, and~\eqref{eq:SQout_integral2} for the configurations (i) and (ii). 
We see that both temperature definitions agree well when the OC efficiently equilibrates the edge channel ($\tau_C T_{\rm in}\ll 1$ for $\tau_C T_{\rm oc}\rightarrow 0$, or $\tau_C T_{\rm oc}\gg 1$ for $\tau_C T_{\rm in}\rightarrow 0$) but clearly deviate when the equilibration is poor and the output channel is out-of-equilibrium due to the heat Coulomb blockade effect. 
Importantly, we therefore find that a measurement of contrasting values for the two effective output temperatures \eqref{eq:T-evaluated}~and~\eqref{eq:T_def_new} is therefore a strong indication that the output channel is out-of-equilibrium.

We can understand the distinctive characteristics of the two effective output temperatures in more detail by analyzing the output bosonic distribution function $n_{\rm out}(\omega)$. To derive this function, we start from Eq.~\eqref{eq:nb_S_relation} which relates the equilibrium charge-current noise $S(\omega)$ [see Eq.~\eqref{eq:S_eq}] and the Bose-Einstein distribution $n_{B}(\omega)$ [see Eq.~\eqref{eq:NB_def}]. By using this relation in the expression for the output charge-current noise ~\eqref{eq:Sout}, we find that
\begin{align}
\label{eq:nBout}
    n_{\rm out}(\omega) &=  \sum_{p=\rm{in},\rm{oc}}|\mathcal{T}_p(\omega)|^2 n_{B,p}(\omega) \notag \\ &= \frac{n_{B,\text{in}}(\omega) +n_{B,\text{oc}}(\omega)\omega^2 \tau_C^2}{1+\omega^2 \tau_C^2},
\end{align}
where $n_{B,p}= (\exp(\omega/T_p)-1)^{-1}$, with $p\in \{\text{in},\text{oc}\}$, are the Bose-Einstein distributions for the density fluctuations (which are bosonic) of the input and OC Langevin source subsystems. Next, we use the output distribution~\eqref{eq:nBout} to re-write the output heat current~\eqref{eq:T_integral} as the following integral over the output bosonic distribution function
\begin{align}
\label{eq:Jout_nout}
J_{\rm out} = \frac{\kappa_0}{2}T^2_{\rm out}= \int_{0}^{\infty} \frac{d\omega}{2\pi} \omega n_{\rm out}(\omega). 
\end{align}
With the bosonization technique (see, e.g., Ref.~\cite{Sukhorukov2016Mar} and Appendix~\ref{sec:Appendix_B}), one may further show (see Appendix~\ref{sec:Appendix_equivalence}) that Eq.~\eqref{eq:Jout_nout} is fully equivalent to an energy integral over the output \textit{electronic} distribution function. From Eq.~\eqref{eq:Jout_nout}, we see that the output temperature $T_{\rm out}$, as previously defined in Eq.~\eqref{eq:T_integral}, is an integrated linear combination of the input and OC Bose-Einstein distributions weighted by $\omega$.

Next, we consider the zero-frequency output heat-current noise, i.e., Eq.~\eqref{eq:SQout} for $\omega=0$. By inserting the relation~\eqref{eq:nb_S_relation} into Eq.~\eqref{eq:SQout} and using the sum rule~\eqref{eq:SumRule} for $|\mathcal{T}_p(\omega)|^2$, as well as the relations $|\mathcal{T}_p(\omega)|^2=|\mathcal{T}_p(-\omega)|^2$ and $1+n_{B,p}(-\omega)=-n_{B,p}(\omega)$, we obtain
\begin{align}
\label{eq:SQ_out_nout}
    S^J_{\rm out}(0) = \kappa_0 \tilde{T}^3_{\rm out} = \frac{1}{2\pi}\int_{0}^{\infty} d\omega \; \omega^2 n_{\rm out}(\omega)\left(1+n_{\rm out}(\omega)\right).
\end{align}
Hence, it is clear that $S^J_{\rm out}(0)$ measures an aspect of the output bosonic distribution $n_{\rm out}(\omega)$ that is distinct from that which $J_{\rm out}$ measures. It is only when $n_{\rm out}(\omega)$ takes the form of an equilbrium Bose-Einstein distribution that the two measures agree and produce $T_{\rm out}=\tilde{T}_{\rm out}$. This happens in three particularly important cases: (i) In global equilibrium, i.e., for $T_{\rm in}=T_{\rm oc}=T$, we have that $T_{\rm out}=\tilde{T}_{\rm out}=T$. (ii) For $|\mathcal{T}_{\rm in}(\omega)|^2=1-|\mathcal{T}_{\rm oc}(\omega)|^2\rightarrow 0$ we have  $T_{\rm out}=\tilde{T}_{\rm out}=T_{\rm oc}$ and (iii) For $|\mathcal{T}_{\rm oc}(\omega)|^2=1-|\mathcal{T}_{\rm in}(\omega)|^2\rightarrow 0$, we obtain $T_{\rm out}=\tilde{T}_{\rm out}=T_{\rm in}$. Cases (ii)-(iii) are in full agreement with the temperature comparisons in Fig.~\ref{fig:temp_comparisons}. 

The above analysis suggests that the two effective output temperatures Eq.~\eqref{eq:T-evaluated} and Eq.~\eqref{eq:T_def_new} can be viewed as defining two distinct ``out-of-equilibrium thermometers'':  
The heat-current based thermometer~\eqref{eq:T-evaluated} targets the average energy, $\omega n_{\rm out}(\omega)$, carried by the output (particle-like) bosonic density fluctuations. The broadening of the average energy can be seen as a temperature measure. This is the temperature that is reached by relaxation with a perfectly coupled floating temperature probe~\cite{Hajiloo2020Oct}. 

In contrast, the heat noise-based thermometer~\eqref{eq:T_def_new} targets deviations from the average energies of the output (particle-like) bosonic density fluctuations, i.e., $\omega^2 n_{\rm out}(\omega) \left(1+n_{\rm out}(\omega)\right)$, which can also be viewed as a temperature measure. In contrast to the average heat current based thermometer, the heat-current noise thermometer is generally a non-linear quantity in the input and OC distributions.

While $T_\mathrm{out}$ and $\tilde{T}_\mathrm{out}$ agree in equilibrium, one might find interesting insights into the edge-channel thermalization process from the way the two temperature approach the equilibrium value, when, e.g., modifying the edge-channel length.

\subsubsection{Impact of potential fluctuations on the heat conductance}
\begin{figure}[t!]
\begin{center}
\captionsetup[subfigure]{position=top,justification=raggedright}
\subfloat[]{
\includegraphics[width=0.95\columnwidth]{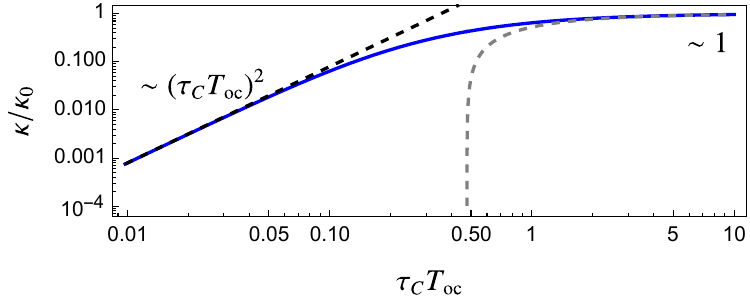}
\label{fig:kappa_renorm_single}}
\\[-0.25cm]
\subfloat[]{
\includegraphics[width =0.95\columnwidth]{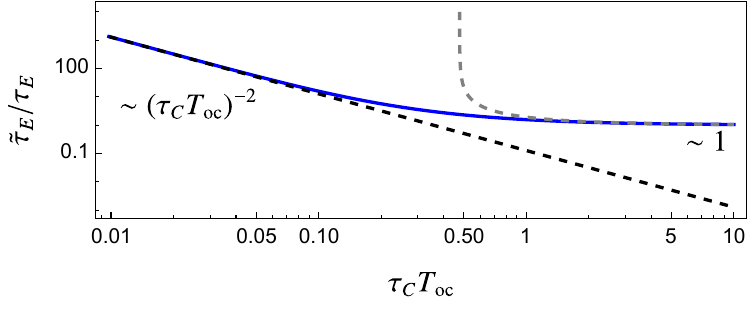}
\label{fig:tauE_renorm_single}}
\caption{ (a) Blue solid line: renormalization of the heat conductance $\kappa T_{\rm oc}/(\kappa_0 T_{\rm oc})$ in Eq.~\eqref{eq:linear_coefficient} vs $\tau_C T_{\rm oc}$, where $\tau_C=R_qC$ is the $RC$ time and $T_{\rm oc}$ is the OC temperature. Black and gray dashed lines are the asymptotic limits in Eq.~\eqref{eq:thermal_cond}.    (b) Blue solid line: Renormalization of the energy relaxation time $\tilde{\tau}_E/\tau_E=\kappa_0/\kappa$ in Eq.~\eqref{eq:New_tau_E} vs $\tau_C T_{\rm oc}$. Black and gray dashed lines are obtained from Eq.~\eqref{eq:thermal_cond}.   }
\label{fig:single_renorm}
\end{center}
\end{figure}

As a next step, we investigate the impact of potential fluctuations on the heat conductance $\kappa T_{\rm oc}$ of the output channel. The heat conductance is an additional relevant observable, in which the effect of the heat Coulomb blockade is experimentally accessible.
At the same time, this quantity is of relevance for the discussion of heat-current noise in the presence of temperature fluctuations, presented below in Sec.~\ref{sec:HeatNoise_Temp_Fluct}. 
Within the linear approximation~\eqref{eq:energyconsfluc}, the first term in the heat-current Langevin equation~\eqref{eq:linearized_total_heat_fluct}, is equivalent to the linear-response coefficient of the output heat current induced by a small change in the OC temperature $T_{\rm oc}(t)$,
\begin{align}
\label{eq:kappa_def}
    \kappa T_{\rm oc} \equiv \frac{\partial J_{\text{out}}}{\partial T_{\rm oc}}   .
\end{align}
 To compute this quantity, we differentiate the output heat current~\eqref{eq:T_integral} with respect to $T_{\rm oc}$ and by using Eq.~\eqref{eq:T-evaluated}, we find 
\begin{align}
\label{eq:linear_coefficient}
    \kappa T_{\rm oc} = \kappa_0 T_{\rm oc}+\frac{1}{4 \pi  T_{\rm oc} \tau_C^2}+\frac{1}{4 \tau_C}-\frac{\psi'\left(\frac{1}{2 \pi  T_{\rm oc} \tau_C}\right)}{8 \pi ^2 T_{\rm oc}^2 \tau_C^3}.
\end{align}
Here, $\psi'(z)$ is the derivative of the function $\psi(z)$ defined below Eq.~\eqref{eq:integral_function}. We remark here that a non-negligible phonon contribution to $\kappa$ would add to the right-hand-side of Eq.~\eqref{eq:linear_coefficient}, but as argued below Eq.~\eqref{eq:energycons}, phonons can be neglected under typical experimental conditions. Equation~\eqref{eq:linear_coefficient} shows that $\kappa$ approaches different asymptotic values depending on the dimensionless parameter $\tau_C T_{\rm oc}$:
\begin{align}\label{eq:thermal_cond}
    \frac{\kappa}{\kappa_0} = \begin{cases}
        \frac{4 \pi ^2}{5}   \tau_C^2 T_{\rm oc}^2& \tau_C T_{\rm oc} \ll 1,\\
        1-\frac{3}{2\pi} \left(\tau_C T_{\rm oc}\right)^{-1} & \tau_C T_{\rm oc} \gg 1.
    \end{cases}
\end{align}
In particular, we see that for $\tau_C T_{\rm oc}\ll 1 $, the heat conductance is strongly suppressed, because the edge channel teleports across the OC, and the particles in the output channel are not influenced by the OC. We plot Eq.~\eqref{eq:linear_coefficient} and~\eqref{eq:thermal_cond} in Fig.~\ref{fig:kappa_renorm_single}.

\subsubsection{Heat-current cross correlations}
Another important consequence of the linearized heat current Langevin equation~\eqref{eq:linearized_total_heat_fluct} is that, in contrast to a conventional Langevin theory, the ``source term'' $\delta J_{\rm out}^{}(t)$ is in fact correlated with the incoming heat-current fluctuations, i.e., 
 \begin{align}\label{eq:SQio}
     \left\langle \delta J_{\rm in}(\omega) \delta J^{}_{\rm out}(\omega') \right\rangle  \equiv 2\pi \delta(\omega+\omega') S^J_{\rm io}(\omega)\neq 0.
 \end{align}
We obtain the heat-current cross-correlations $ S^J_{\rm io}(\omega)$ with the same approach as for the output noise Eq.~\eqref{eq:SQout}, with the result
\begin{align}
\label{eq:Sio}
     S_{\rm io}^{J}(\omega) &= \frac{R_q^2}{4\pi}\int_{-\infty}^{\infty}d\omega_1 |\mathcal{T}_{\rm in}(\omega_1)|^2 S_{\rm in}(\omega_1)  S_{\rm in}(\omega-\omega_1).
\end{align} 
In the zero-frequency limit, this expression reduces to the dimensionless integral expression 
\begin{align}\label{eq:io_asym}
    S_{\rm io}^J(0)  = \frac{T_{\rm in}^3}{4\pi}\int_{-\infty}^{\infty }dz \frac{z^2}{(e^z-1)(1-e^{-z})} \frac{1}{(\left(\tau_C T_{\rm in}\right)^2 z^2+1)}, 
\end{align}
with the asymptotic limits comparing to $S^J_{\rm in}(0)=\kappa_0T_{\rm in}^3$,
\begin{align}
\label{eq:io_asym2}
    \frac{S_{\rm io}^J(0)}{S^J_{\rm in}(0)}
    \approx\begin{cases}
   1 - \frac{4\pi^2}{5}(\tau_C T_{\rm in})^2 &    \tau_C  T_{\rm in} \ll 1,\\
\frac{3}{2\pi} \left(\tau_C T_{\rm in}\right)^{-1} & \tau_C T_{\rm in} \gg 1.
\end{cases}
\end{align}
Hence, also these nonvanishing correlations between input and output heat current fluctuations are affected by the heat Coulomb blockade.

We end this subsection by emphasizing that its results only hold in the limit of negligible OC temperature fluctuations $\Delta T_{\rm oc}(t)=0$, which amounts to assuming an infinite heat capacity, $C_E[T_{\rm oc}(t)] \propto \tau_{E}\rightarrow \infty$. It is however known~\cite{Dashti2018Aug} that heat-current fluctuations are highly sensitive to sizeable temperature fluctuations and energy relaxation times. It is therefore natural to ask which additional features in the heat dynamics that emerge in the presence of finite OC heat capacity and OC temperature fluctuations. This is the question to which we turn next.  

\subsection{Heat dynamics in the presence of temperature fluctuations}
\label{sec:HeatNoise_Temp_Fluct}
We now analyze how the output heat-current noise is impacted by sizeable fluctuations of the OC temperature, $T_{\rm oc}(t) = T_{\rm oc} +\Delta T_{\rm oc}(t)$. Their origin in the setup in Fig.~\ref{fig:Single_Channel_Setup} is due to a finite OC heat capacity $C_E[T_{\rm oc}(t)]$. 

In general, the \textit{average} temperature in the floating, energy-conserving, OC, $T_{\rm oc}$ is given by a power-balance equation based on Eq.~\eqref{eq:energycons} for the average energy currents. 
This average temperature depends on the type of input to the OC and on possible coupling to further heat baths, e.g., a phonon bath, see Ref.~\cite{Dashti2018Aug}. In the simple situation considered in the present section and indicated in Fig.~\ref{fig:Single_Channel_Setup}, we will always find $T_{\rm oc}=T_{\rm in}$.

In order to obtain the fluctuations of the heat current as a result of both potential \textit{and} temperature fluctuations, we follow analogous steps to the calculation of the charge-current fluctuations in Sec.~\ref{sec:Single_channel_charge_sector}. More concretely, using a Fourier transform, we solve Eqs.~\eqref{eq:energycons} and~\eqref{eq:heatfluct} for the total output heat-current fluctuations $\delta J_{\rm out}^{\rm tot}(\omega)$, i.e., including also temperature fluctuations. The solution reads~\footnote{The linear coefficients $\mathcal{T}^J_{p}(\omega)$ for $p\in \{\text{in},\text{out}\}$ in Eq.~\eqref{eq:Jtot_langevin}, in contrast to their charge counterparts~\eqref{eq:t_values}, cannot be viewed as scattering amplitudes of the heat-current fluctuations. The similarity to Eq.~\eqref{eq:t_values} is due to the fact that the equations for the heat-current fluctuations are linearized.}
\begin{subequations}    
\label{eq:Jtot_langevin}
\begin{align}
    &\delta J_{\rm out}^\mathrm{tot}(\omega) = \mathcal{T}^J_{\rm in}(\omega) \delta J_{\rm in}(\omega) + \mathcal{T}^J_{\rm out}(\omega) \delta J^{}_{\rm out}(\omega),\label{eq:deltaJSol}\\
    & \mathcal{T}^J_{\rm in}(\omega) = 1-\mathcal{T}^J_{\rm out}(\omega) = \left[1-i\omega \tilde{\tau}_E\right]^{-1}\label{eq:tq_values}.
\end{align}
We will in the following discuss the relevant ingredients to these fluctuations occurring at different time scales as well as the resulting heat-current noise. 
\end{subequations}

\subsubsection{Fast heat-current fluctuations}
Under the assumption of the separation of time scales~\eqref{eq:hierachy_time}, the OC temperature fluctuations can be viewed as slow fluctuations (with characteristic timescale $\tau_E$) superimposed on fast heat-current fluctuations induced by charge-current fluctuations (with time scale  $\tau_C \ll \tau_E$). This is captured by the heat dynamics equations~\eqref{eq:heatfluct}-\eqref{eq:linearized_total_heat_fluct}. The starting point for our analysis of the total (i.e., including also temperature fluctuations) output heat-current noise is the second, ``source'', term in Eq.~\eqref{eq:linearized_total_heat_fluct} given as
\begin{align}
\label{eq:heat_source}
    \delta J_{\text{out}}^{}(t) \equiv \frac{R_q}{2}\left[(\delta I_{\rm out}(t))^2 - \langle (\delta I_{\rm out}(t))^2 \rangle\right].
\end{align}
This term takes the role of a semi-classical source of heat-current fluctuations due to charge-current fluctuations. The noise corresponding to these fast heat-current fluctuations $\delta J_{\rm out}^{}(t)$ is given by the previously calculated heat-current noise of the OC subject only to potential fluctuations~\eqref{eq:SQout}, i.e., 
\begin{align}
    \left\langle \left\langle \delta J_{\rm out}^{} (\omega) \delta J_{\rm out}^{} (\omega')  \right\rangle \right\rangle_E \equiv 2\pi \delta(\omega+\omega') S^J_{\rm out}(\omega).
\end{align}
This identification connects the Langevin approaches for charge and heat currents and ensures consistency of our approach. 

\subsubsection{Enhancement of the energy relaxation time}
From the influence of the $RC$ time $\tau_C$ on the heat conductance $\kappa T_{\rm oc}$ in Eqs.~\eqref{eq:linear_coefficient} and~\eqref{eq:thermal_cond}, we see that in the heat Coulomb blockade regime, a new time scale emerges,
\begin{align}
\label{eq:New_tau_E}
\tilde{\tau}_E &\equiv \frac{C_E[T_{\rm oc}]}{\kappa T_{\rm oc}} \nonumber\\
&\sim \frac{\tau_E}{T_{\rm oc}^2 \tau_C^2} \gg \tau_E,\quad \text{for\ } \tau_C T_{\rm oc} \ll 1.
\end{align}
 This time scale enters in Eq.~\eqref{eq:tq_values}. In physical implementations as considered here, where time scales of charge and energy dynamics are separated~\eqref{eq:hierachy_time}, it is much larger than the energy-relaxation time $\tau_E$. Its dependence on the $RC$-time is shown in Fig.~\ref{fig:tauE_renorm_single}. Equation~\eqref{eq:New_tau_E} hence implies that in the heat Coulomb blockade regime, namely for $ \tau_C T_{\rm oc} \ll 1$, the OC temperature fluctuations, induced by the finite $C_E[T_{\rm oc}]$, are strongly suppressed, as shown in Fig.~\ref{fig:tauE_renorm_single}. The consequences of this feature for the dynamics of temperature fluctuations are further shown  in Eq.~\eqref{eq:temp_supp} below. 

\subsubsection{Total heat-current noise impacted by potential- and temperature fluctuations}
From Eqs.~\eqref{eq:deltaJSol} and~\eqref{eq:tq_values}, we obtain the total output heat-current noise in the presence of temperature fluctuations, denoted as
\begin{align}
\label{eq:chi_out}
   S^{J,\text{tot}}_{\rm out}(\omega) \equiv  2\pi \delta (\omega+\omega')\left\langle\left\langle \delta J^{\rm tot}_{\rm out}(\omega) J^{\rm tot}_{\rm out}(\omega') \right\rangle \right\rangle_E.
\end{align}
The separation of time scales, $\tilde{\tau}_E^{-1} \ll T_{\rm in},T_{\rm oc}$, implies that the contributions from all faster heat-current fluctuations, i.e., $S^J_{\rm in}(\omega)$, $ S_{\rm io}^{J}(\omega)$, and $S^J_{\rm out}(\omega)$, are well captured in $S^{J,\text{tot}}_{\rm out}(\omega)$ by their zero-frequency contributions, only.
Therefore, by inserting the solutions~\eqref{eq:deltaJSol} and~\eqref{eq:tq_values} into \eqref{eq:chi_out}, we find 
\begin{align}
\label{eq:SoutQTfluc2}
   S^{J,\text{tot}}_{\rm out}(\omega) =  \sum_{p=\rm{in},\rm{out}}|\mathcal{T}^J_p(\omega)|^2 S^J_p(0).
\end{align}  
Interestingly, the heat-current cross-correlations $ S_{\rm io}^{J}(\omega)$, defined in Eq.~\eqref{eq:SQio}, do not enter in Eq.~\eqref{eq:SoutQTfluc2}, due to a cancellation by the identity $\mathcal{T}_{\rm in}^J(\omega) \mathcal{T}_{\rm out}^J(-\omega) = - \mathcal{T}_{\rm out}^J(\omega) \mathcal{T}_{\rm in}^J(-\omega)$. Instead, the absolute values of $\mathcal{T}^J_p$ enter Eq.~\eqref{eq:SoutQTfluc2}. They fulfill a sum rule
\begin{align}\label{eq:heat_sumrule}
\sum_{p=\rm{in},\rm{out}}|\mathcal{T}^J_p(\omega)|^2 =1,
\end{align}
similar to the charge dynamics, but which here reflects energy conservation.

We emphasize that the expressions~\eqref{eq:Jtot_langevin}-\eqref{eq:SoutQTfluc2} hold in a limit where we consider OC potential fluctuations exactly, but treat the OC temperature fluctuations to linear order.
Therefore, Eq.~\eqref{eq:SoutQTfluc2} extends the result in Ref.~\cite{Dashti2018Aug} by incorporating the additional effects of OC potential fluctuations into the modified energy relaxation time $\tilde{\tau}_E$. 

\begin{figure}[t!]
\includegraphics[width =0.99\columnwidth]{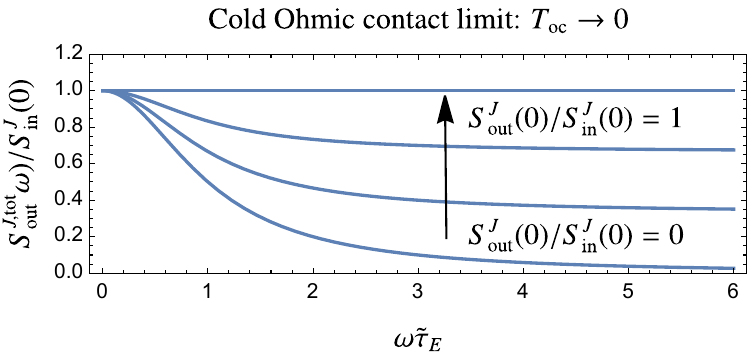}
\caption{Total output heat current noise $S^{J,\text{tot}}_{\rm out}(\omega)$ in Eq.~\eqref{eq:SJ_tot_single} vs $\omega \tilde{\tau}_E$, where $\omega$ is the measurement frequency, and $\tilde{\tau}_E$ is the enhanced energy relaxation time~\eqref{eq:New_tau_E}. The different curves depict different values of the ratio $S^{J}_{\rm out}(0)/S^{J}_{\rm in}(0)$, in the limit $T_{\rm oc}/T_{\rm in}\rightarrow 0$, ranging from $0$ (bottom curve) to 1 (top curve) in steps of 1/3.}
\label{fig:STot_single_interpolation}
\end{figure}

From the results Eqs.~\eqref{eq:kappa_def}-\eqref{eq:New_tau_E}, we are now able to compute the total output heat-current noise in the presence of both OC potential- and OC temperature fluctuations.  
We find
\begin{subequations}
\label{eq:SJ_tot_single}
\begin{align}
    S^{J,\text{tot}}_{\rm out}(\omega) & = \frac{S_{\rm in}^J(0)+\omega^2\tilde{\tau}_E^2 S_{\rm out}^J(0)}{1+\omega^2\tilde{\tau}_E^2}\\
   & \rightarrow  \left\{\begin{array}{ll}
   S^{J}_{\rm in}(0)  & \ \omega\tau_C\ll\omega\tilde{\tau}_E \ll 1,\\
   S^{J}_{\rm out}(0)  &  \  \omega\tau_C \ll 1 \ll\omega\tilde{\tau}_E,
   \end{array}\right.
   \label{eq:SQtot_cases}
  \end{align}
\end{subequations}
where $S^{J}_{\rm in}(0)$ and $S^{J}_{\rm out}(0)$ are the input and output low-frequency heat-current fluctuations in the absence of temperature fluctuations, respectively. Equation~\eqref{eq:SJ_tot_single} is a key result of our work, showing that the modified time scale of the energy dynamics induces a further band-pass filtering effect, impacting the heat-current fluctuations. This new band-pass filtering effect is formally analogous to the one underlying the heat Coulomb blockade, but is due to a physically distinct effect, namely it is induced by a finite \textit{heat} capacity. At frequencies far below the inverse of the modified energy relaxation time, the total output heat-current fluctuations, $S^{J,\text{tot}}_{\rm out}(\omega)$,  are uniquely given by the low frequency input heat-current fluctuations $S^{J}_{\rm in}(0)$. Hence, the equilibration by the OC is impeded.
This is a consequence of energy conservation and the floating nature of the OC. 
In other words, at sufficiently low measurement frequency, the averaged heat-current fluctuations do not ``see'' effects of the OC heat capacity. An important insight from this feature is that the OC temperature fluctuations can result in an averaging effect that fully masks the heat-current fluctuations induced by potential fluctuations, which would in the limit of $\tau_E\to\infty$ always be visible, even in the low-frequency output heat-current noise, see Eq.~\eqref{eq:SQout}. Indeed, with growing frequency, the fast output heat-current fluctuations induced by the OC potential fluctuations increasingly contribute to the total output heat-current noise. 
This can be seen in Fig.~\ref{fig:STot_single_interpolation}, where we, for concreteness, plot  Eq.~\eqref{eq:SJ_tot_single} in the configuration of a very cold OC. In this case, the ratio $S^{J}_{\rm out}(0)/S^{J}_{\rm in}(0)$ ranges between 0 and 1, see Eq.~\eqref{eq:SQout_asymptotic_1}, depending on the parameter $\tau_C T_{\rm in}$. 

We remark that for the single-channel setup considered here, the experimentally required measurement frequency might be hard to reach, due to the strong renormalizaton of the energy-relaxation time, see Eq.~\eqref{eq:New_tau_E} and Fig.~\ref{fig:tauE_renorm_single}. Particular care must be taken when $\tilde{\tau}_E$ approaches $\tau_{\rm e-ph}$, see Eq.~\eqref{eq:hierachy_time}.
The multi-channel setup, where the renormalization is weaker, see Eq.~\eqref{eq:tau_E_blocked} below, might facilitate  reaching low enough frequency.

\subsubsection{OC temperature fluctuations}
We finally investigate the behaviour of the OC temperature fluctuations due to the emergence of the new energy relaxation time~\eqref{eq:New_tau_E} in the Coulomb blockade regime. To this end, we analyze the OC temperature-fluctuations correlation function, defined as 
 \begin{align}
     \left\langle \left\langle 
    \Delta T_{\rm oc} (\omega) \Delta T_{\rm oc} (\omega') \right\rangle \right\rangle_E \equiv 2\pi \delta (\omega+\omega') S^{T}_{\rm oc}(\omega). 
 \end{align}
We obtain this function by solving Eqs.~\eqref{eq:energycons},~\eqref{eq:energyconsfluc}, and~\eqref{eq:deltaJSol} and  obtain
\begin{align}\label{eq:ST_gen}
 S^T_{\rm oc}(\omega) =\frac{S_{\rm in}^J(0) - 2 S_{\rm io}^J(0)+S_{\rm out}^J(0)}{\kappa^2 T_{\rm oc}^2\left(1+\omega^2 \tilde{\tau}_E^2\right)},
\end{align}
where $S_{\rm in}^J(0)=\kappa_0 T_{\rm in}^3$, $S_{\rm out}^J(0)$ is obtained from Eq.~\eqref{eq:SQout} by taking $\omega=0$, and $S_{\rm io}^J(0)$ is given in Eqs.~\eqref{eq:io_asym} and~\eqref{eq:io_asym2}. The relation~\eqref{eq:ST_gen} establishes a connection between heat \textit{current} fluctuations and \textit{local} temperature fluctuations.
In the simple single-channel setting with a thermal input, where we have $T_{\rm in} = T_{\rm oc}=T$ for the average temperatures, Eq.~\eqref{eq:ST_gen} evaluates to 
\begin{align}
\label{eq:ST_equal_T}
 S^T_{\rm oc}(\omega)=  
   \frac{2 T   }{\kappa \left(1+\omega^2 \tilde{\tau}_E^2\right)}.
\end{align}
We see that this takes the same functional form regardless of whether the OC is in the heat Coulomb blockade regime or not. This result is related to the fact that in equilibrium, we have $S_{\rm in}(\omega) = S_{\rm oc}(\omega) = S(\omega)$, which implies that the numerator in Eq.~\eqref{eq:ST_gen} simplifies as
\begin{multline}
\label{eq:temp_identity}
    S_{\rm in}^J(0) - 2 S_{\rm io}^J(0)+S_{\rm out}^J(0) \\= R_q^2 \int_{-\infty}^{\infty}\frac{d\omega'}{2\pi} \left(1-\left|\mathcal{T}_{\rm in}(\omega')\right|^2\right) S(\omega') S(-\omega') \\=  T^2 \partial_T R_q \int_{-\infty}^{\infty}\frac{d\omega'}{2\pi} \left|\mathcal{T}_{\rm oc}(\omega')\right|^2  S(\omega) = 2 \kappa T^3.
\end{multline}
Here, in the first equality, we used Eqs.~\eqref{eq:SQout} and~\eqref{eq:io_asym}, and in the second equality, we used the sum rule~\eqref{eq:SumRule}. In the time-domain, Eq.~\eqref{eq:ST_equal_T} corresponds to an exponential suppression of the temperature correlations
\begin{align} \label{eq:temp_supp}
    \left\langle \left\langle 
    \Delta T_{\rm oc} (t) \Delta T_{\rm oc} (0) \right\rangle \right\rangle_E & = \left\langle \left\langle 
    \Delta T_{} (t) \Delta T_{} (0) \right\rangle \right\rangle_E \notag \\ &=   \frac{T^2}{C_E[T]} e^{-|t|/\tilde{\tau}_E}.
\end{align}
While the amplitude of the temperature-fluctuations correlation function~\eqref{eq:temp_supp} is the universal, expected result (see, e.g., Ref.~\cite{Pekola21Oct}), its decay rate, $\tilde{\tau}_E$, is in fact sensitive to $\tau_C$ via Eqs.~\eqref{eq:thermal_cond} and~\eqref{eq:New_tau_E}, even in the global equilibrium case. In other words, as a result of heat Coulomb blockade, for temperatures $\tau_C T\ll 1$, the temperature-fluctuations correlations decay much slower than for $\tau_C T\gg 1$.

\section{Generalization to multiple edge channels}
\label{sec:Multi_Gen}
\begin{figure}[bt]
\includegraphics[width =.9\columnwidth]{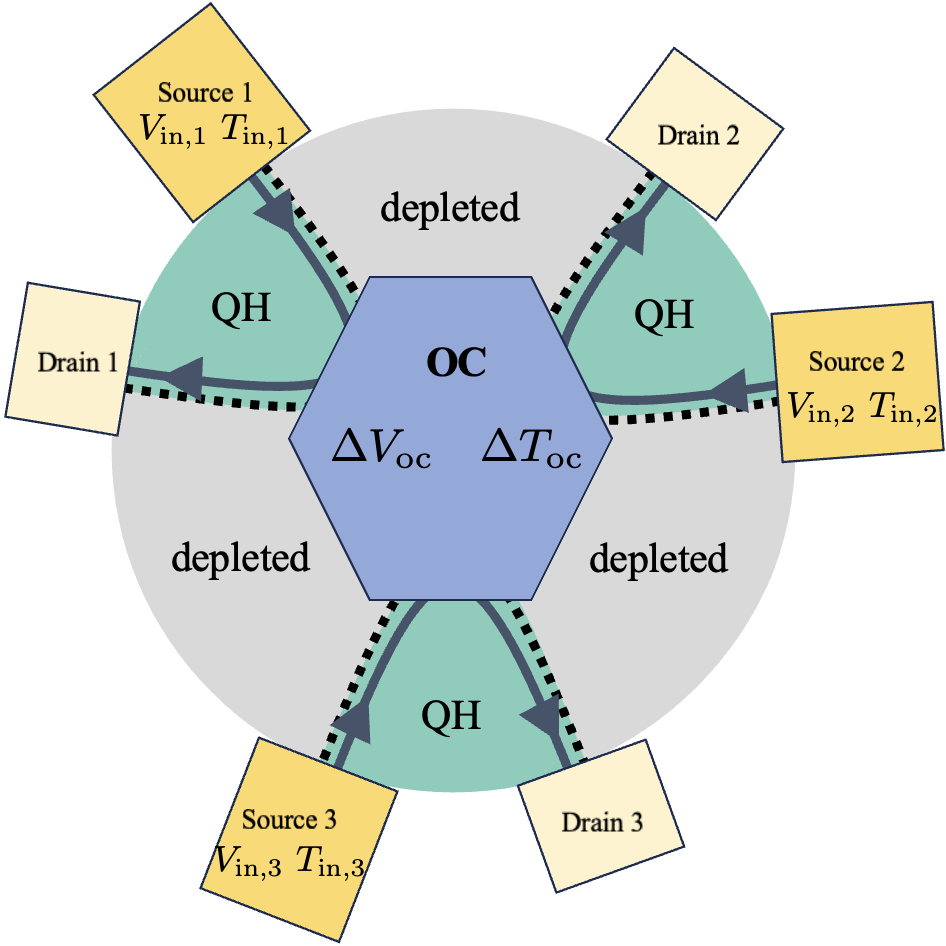}
\caption{Schematics of an Ohmic contact (OC, blue) connected to multiple (here $N=3$), chiral edge channels in the quantum Hall (QH) regime.  Pairs of incoming/outgoing edge channels appear on the boundaries (black, dotted lines) between QH bulk regions (turquoise regions, filling factor $\nu$) and depleted regions (gray regions). All input channels are fed from large metallic contacts (in yellow) characterized by input voltages, $V_{\rm in}$, and input temperatures, $T_{\rm in}$, for $m=1,\hdots,N$. Fluctuations carried by the input edge channels generate potential, $\Delta V_{\rm oc}$, and temperature, $\Delta T_{\rm oc}$, fluctuations. In turn, these quantities produce charge- and heat-current fluctuations in the outgoing channels, which can be detected in the drain contacts (in light yellow). }
\label{fig:Multi_Channel_Setup}
\end{figure}

In the previous sections, we focused on the simplest setting of a single edge channel connected to the OC, see Fig.~\ref{fig:Single_Channel_Setup}. 
Here, we generalize this description to the situation with $N$ incoming and $N$ outgoing edge channels, as depicted in Fig.~\ref{fig:Multi_Channel_Setup}. 
Such multi-channel setups have been used extensively to extract the edge heat conductance of a broad range of QH states~\cite{Jezouin2013Nov,Banerjee2017May,Banerjee2018Jul,Srivastav2019Jul,Srivastav2021May,Melcer2022Jan,Srivastav2022Sep,LeBreton2022Sep,Dutta2022Sep}. 
For simplicity, we assume in the following that all connected edge states are identical, by bordering bulk regions with the same filling factors, chosen here as $\nu=1$.
This means that they have equal charge conductances, $R_q^{-1}=e^2/2\pi$, and equal heat conductance prefactors $\kappa_0$. Technically, we take the channels to have equal ``central charges'' $c=1$, see, e.g., Ref.~\cite{francesco2012conformal}. Our approach can however be straightforwardly extended to setups with composite edge states, including non-Abelian edge channels with $c\neq 1$ (see, e.g., Refs.~\cite{Banerjee2018Jul,Park2020Oct,Dutta2022Sep,Hein2023Jun} for descriptions of such setups).

In comparison to the single-channel setup, several novel features emerge in the multi-channel setup. With our $N$-channel charge dynamics equations~\eqref{eq:Charge_dyn_N}, we confirm the heat Coulomb blockade effect~\cite{Sivre2018Feb,sivre2019electrical} which in the $N$-channel setup is manifest by a reduction of the total heat conductance by \textit{one unit} of $\kappa_0 T_{\rm oc}$ [see Eqs.~\eqref{eq:thermal_cond_Ncase} and~\eqref{eq:kappa_tot_reduced} below]. We further show in Eq.~\eqref{eq:tau_E_blocked} below how this reduction impacts the $N$-channel OC energy relaxation time. Finally, we also show how combinations of auto- and cross correlations can be used to extract the OC potential- and OC temperature fluctuations, which is a highly non-trivial task in the single-channel case.

\subsection{Multi-channel Langevin approach}\label{sec:Multichannel_approach}
In order to proceed, we generalize the Langevin approach presented in Sec.~\ref{sec:BL_approach} to the multi-channel situation as sketched in Fig.~\ref{fig:Multi_Channel_Setup}. For the charge dynamics in the $N$-channel case, the Langevin equations~\eqref{eq:Kirchhoff} and~\eqref{eq:Langevin} generalize  to
\begin{subequations}
    \begin{align}
    \frac{dQ(t)}{dt} &= \sum_{m=1}^N \left[ I_{\text{in},m}(t) - I_{\text{out},m}(t)\right],\label{eq:Kirchhoff_Ncase}\\
    \delta I_{\text{out},m}(t) &= \frac{\Delta Q(t)}{N \tau_{C,N}}  + \delta I_{\text{oc},m}(t).\label{eq:Langevin_Ncase}
\end{align}
\end{subequations}
Here, $\Delta Q(t)$ are the temporal OC charge fluctuations and $ I_{\text{in},m}(t)$, $I_{\text{out},m}(t)$, and $I_{\text{oc},m}(t)$  are the incoming, outgoing, and thermally induced charge currents on branch $m$, respectively, with $\delta...$ denoting the corresponding fluctuations. Here, ``branch'' refers to one pair of input and output channels, see Fig.~\ref{fig:Multi_Channel_Setup}. Moreover, $\tau_{C,N}\equiv\tau_{C}/N$ denotes the \textit{total} $RC$ time of the OC when $N$ channels are attached. 

For the $N$-channel heat dynamics, Eqs.~\eqref{eq:energycons} and~\eqref{eq:linearized_total_heat_fluct} generalize to 
\begin{subequations}   
\label{eq:NChannel_heat_Dynamics}
 \begin{align} \label{eq:energycons_Ncase}
    & \frac{dU(t)}{dt} = \sum_{m=1}^N \left[ J^{\rm tot}_{\text{in},m}(t) - J^{\rm tot}_{\text{out},m}(t)\right],\\
     &\delta J^{\rm tot}_{\text{out},m}(t) = \kappa_m T_{\rm oc} \Delta T_{\rm oc}(t) + \delta J^{}_{\text{out},m}(t)
     \label{eq:heatfluct_Ncase},\\
    & \delta J^{}_{\text{out},m}(t) \equiv \frac{R_q}{2}\left[(\delta I_{\text{out},m}(t))^2 - \langle (\delta I_{\text{out},m}(t))^2 \rangle\right].\label{eq:Jfast_Ncase}
\end{align}
\end{subequations}
Here, $U(t)$ is the internal OC energy,  $ J^{\rm tot}_{\text{in/out},m}(t)$ is the total incoming/outgoing time-dependent heat current on branch $m$, with $\delta...$ denoting the corresponding fluctuations, and $\Delta T_{\rm oc}(t)$  are the OC temperature fluctuations. The explicit form of the heat conductance $\kappa_m T_{\rm oc}$ will be discussed later, see Eq.~\eqref{eq:J_out_m}. Similarly to the single channel case [see Eq.~\eqref{eq:heat_source}], the heat ``source'' term~\eqref{eq:Jfast_Ncase} for branch $m$ is given in terms of the charge-current fluctuations on the same branch, via Eq.~\eqref{eq:JI_relation}.  

\subsection{Multi-channel charge dynamics}
\label{sec:Multi_Gen_Charge}
We start by considering charge current fluctuations in the ``standard" heat Coulomb blockade regime, namely where temperature fluctuations can be neglected.

\subsubsection{Charge-current auto- and cross-correlations in the presence of OC potential fluctuations}
Analogously to the single-channel case, see Sec.~\ref{sec:Single_channel_charge_sector}, we solve Eqs.~\eqref{eq:Kirchhoff_Ncase} and~\eqref{eq:Langevin_Ncase} for the output average charge currents and  charge-current fluctuations,
\begin{subequations}
\label{eq:Charge_dyn_N}
\begin{align}
    &I_{\text{out},m}(\omega) = \sum_{p=\rm{in},\rm{oc}}\sum_{n=1}^N\mathcal{T}_{p,mn}(\omega) I_{p,n}(\omega),\label{eq:jSol_Ncase}\\
    &\delta I_{\text{out},m}(\omega) = \sum_{p=\rm{in},\rm{oc}}\sum_{n=1}^N\mathcal{T}_{p,mn}(\omega)\delta I_{p,n}(\omega),
\end{align}
with the coefficients
\begin{align}
    &\mathcal{T}_{{\rm oc},mn}(\omega) \equiv  \delta_{mn}  - N^{-1}\left[1-i\omega  \tau_{C,N} \right]^{-1},
    \\
    & \mathcal{T}_{{\rm in},mn}(\omega) \equiv  N^{-1}\left[1-i\omega  \tau_{C,N} \right]^{-1}
    \label{eq:t_values_Ncase},
    \end{align}
 and find the output charge-current noise 
 \begin{align}
    &   S_{\text{out},m}(\omega) = \sum_{p=\rm{in},\rm{oc}} \sum_{n=1}^N   \left|\mathcal{T}_{p,mn}(\omega) \right|^2  S_{p,n}(\omega).\label{eq:Sout_NCase}
\end{align}
\end{subequations}
For simplicity we take, here and below, the input charge-current noises from all branches to be equal, $S_{\text{in},m}(\omega)=S_{\rm in}(\omega)$, by choosing all source contacts to have the same temperature, $T_{\text{in},m}=T_{\rm in}$. In addition, we choose $V_{\text{in},m}=0$; as a consequence, also the average OC potential vanishes, $V_{\rm oc}=0$. We furthermore use the fact that the thermal OC charge-current fluctuation contributions to each output channel are equal $S_{\text{oc},m}(\omega)=S_{\rm oc}(\omega)$. With these choices, spelling out Eq.~\eqref{eq:Sout_NCase}, results in \begin{align}
\label{eq:Sout_Ncase_2}
    S_{\text{out},m}(\omega) =\frac{S_{\rm in}(\omega)+S_{\rm oc}(\omega)\left[N-1+N\omega^2\tau_{C,N}^2\right]}{N(1+\omega^2\tau_{C,N}^2)}.
\end{align}
This means that the charge-current noise is independent of the specific output channel $m$, but depends on the total number of channels $N$. In the zero-frequency limit, $\omega \tau_{C,N}\ll 1$, Eq.~\eqref{eq:Sout_Ncase_2} reduces to
\begin{align}
\label{eq:Sout_Ncase_3}
    S_{\text{out},m}(0) &=\frac{S_{\rm in}(0)+S_{\rm oc}(0)(N-1)}{N} \notag \\
    &= \frac{R_q^{-1}\left(T_{\rm in} + T_{\rm oc}(N-1)\right)}{N}.
\end{align}
As expected, Eq.~\eqref{eq:Sout_Ncase_3} reduces to Eq.~\eqref{eq:ZeroFreqTemp} for $N=1$. 
However, Eq.~\eqref{eq:Sout_Ncase_3} shows that a measurement of the low-frequency output noise in the multi-channel setup with $N>1$ gives access to the OC temperature $T_{\rm oc}$. The low-frequency noise is often used in the process of extracting the heat conductance of edge channels, see, e.g., Ref.~\cite{Melcer2022Jan} for a detailed analysis. 

What has been much less considered in this context is that the $N$-channel setup introduces cross-correlations between current fluctuations of different output channels. For $m\neq n$,  we define the charge-current cross-correlations as
\begin{align}
\label{eq:S_cross_def}
    \langle \delta I_{\text{out},m}(\omega) \delta I_{\text{out},n}(\omega') \rangle \equiv 2\pi \delta(\omega+\omega')S_{\text{out},m,n}(\omega),
\end{align}
evaluating to
\begin{align}
\label{eq:S_cross_explicit}
    S_{\text{out},m,n} (\omega) = \frac{S_{\rm in}(\omega)-S_{\rm oc}(\omega)}{N(1+\omega^2\tau_{C,N}^2)}, \quad N>1.
\end{align}
As expected, for global equilibrium conditions $T_{\rm in}=T_{\rm oc}$, the cross-correlations vanish. They are also irrelevant with respect to the auto-correlation difference $S_{\rm in}(\omega)-S_{\rm oc}(\omega)$ on frequency- (respectively time-) scales, for which $\omega\tau_{C,N}$ is very large. Equation~\eqref{eq:S_cross_explicit} for the finite-frequency charge-current cross-correlations can be connected to the auto-correlations by comparing with the charge-current noises at finite frequency, \eqref{eq:Sout_Ncase_2} and~\eqref{eq:S_cross_explicit}. We obtain the relation
\begin{align}\label{eq:relation_corr}
    S_{\text{out},m,n} (\omega)
    = S_{{\rm out},m}(\omega) -S_{\rm oc}(\omega), \quad N>1.
\end{align}
In the low-frequency limit, $\omega\tau_{C,N}\ll 1$, and when the OC temperature and the temperatures of the source contacts differ,  $S_{\text{out},m,n} (\omega)$ reduces to
\begin{align}
    \label{eq:S_cross_explicit_zero_freq}
    S_{\text{out},m,n} (0) = \frac{R_q^{-1}\left(T_{\rm in}-T_{\rm oc}\right)}{N}, \quad N>1.
\end{align}
By combining the expressions for the zero-frequency output auto- and cross-correlations, Eq.~\eqref{eq:Sout_Ncase_3} and~\eqref{eq:S_cross_explicit_zero_freq}, we obtain the sum rule
\begin{align}\label{eq:sumrule_Ncase_zerofreq}
    & \sum_{m=1}^N S_{\text{out},m} (0) + \sum_{n=1,m\neq n}^N S_{\text{out},m,n} (0) = N S_{\rm in}(0),
\end{align}
which reflects charge conservation in the $N$-channel setup. 

Interestingly, charge-current auto- and cross-correlations in multi-terminal systems also give access to the OC potential fluctuations: By combining Eq.~\eqref{eq:relation_corr} with the $N$-channel Langevin equation~\eqref{eq:Langevin_Ncase}, we extract the OC potential-fluctuations correlation function, $\left\langle \Delta V_{\rm oc}(\omega)  \Delta V_{\rm oc}(\omega')\right\rangle$, as a combination of output noises,
\begin{multline}
    \left\langle \Delta V_{\rm oc}(\omega) \Delta V_{\rm oc}(\omega')\right\rangle = 2\pi \delta(\omega+\omega') \\ \times \frac{R_{q}^2}{N^2} \frac{S_{\rm in}(\omega) + S_{{\rm out},m}(\omega)-S_{{\rm out},m,n}(\omega)}{N(1+\omega^2 \tau_{C,N}^2)} ,\ \text{for\ } N>1.
\end{multline} 
We thus see that a combination of noise measurements gives access to the OC potential-fluctuations correlation function. This possibility should be contrasted with the single-channel setup, $N=1$, where the OC potential-fluctuations correlations  can not be straightforwardly separated from the thermally induced charge-current fluctuations $\delta I_{\rm oc}$ coming from the OC. Indeed, solving Eq.~\eqref{eq:Kirchhoff} and Eq.~\eqref{eq:Langevin} for $\Delta V_{\rm oc}(\omega)=\Delta Q(\omega)/C$, we find the correlation function for the single-channel case
\begin{align}
    \left\langle \Delta V_{\rm oc}(\omega) \Delta V_{\rm oc}(\omega')\right\rangle & = 2\pi \delta(\omega+\omega') \notag \\ & \times R_{q}^2 \frac{S_{\rm in}(\omega) + S_{\rm oc}(\omega)}{(1+\omega^2 \tau_{C}^2)},\ \text{for\ } N=1, 
\end{align} 
so that both $S_{\rm in}(\omega)$ and $S_{\rm oc}(\omega)$ are needed to fully classify the OC potential correlations in this single-channel case.
\subsubsection{Heat-current auto- and cross correlations in the presence of OC potential fluctuations}
\label{sec:heatnoise_voltage_N}
Going beyond the charge-current noise, we now compute the output heat-current noise in the presence of potential fluctuations, while still neglecting temperature fluctuations. As for the charge dynamics presented in the previous section, we find that there are heat-current auto-correlations, $S^J_{\text{out},m}(\omega)$, as well as heat-current cross-correlations $S^J_{\text{out},m,n}(\omega)$ with $m\neq n$,
\begin{subequations}
\label{eq:Sio_def}
    \begin{align}
        & \left\langle  \delta J_{\text{out},m}^{} (\omega) \delta J_{\text{out},m}^{} (\omega')\right\rangle \equiv 2\pi  \delta(\omega+\omega') S_{\text{out},m}^J(\omega), \\
        & \left\langle  \delta J_{\text{out},m}^{} (\omega) \delta J_{\text{out},n}^{} (\omega')\right\rangle \equiv 2\pi  \delta(\omega+\omega') S_{\text{out},m,n}^J(\omega),\label{eq:SJmn}
    \end{align}
\end{subequations}
with the heat current fluctuations $\delta J_{\text{out},m}^{} (\omega)$ from Eq.~\eqref{eq:Jfast_Ncase}. 

The identity~\eqref{eq:JI_relation} allows use to compute the heat current noises~\eqref{eq:Sio_def} by inserting Eqs.~\eqref{eq:Sout_Ncase_2}, respectively Eq.~\eqref{eq:S_cross_explicit}, into the convolution formula~\eqref{eq:SQ}. This results in
\begin{subequations}
\label{eq:SQ_out_combined}
\begin{align}
\label{eq:SQm_out}
     &S_{\text{out},m}^{J}(\omega) = \frac{R_q^2}{4\pi}\int_{-\infty}^{\infty}d\omega_1 S_{\text{out},m}^{}(\omega_1)  S_{\text{out},m}^{}(\omega-\omega_1), \\
\label{eq:crossheat}
  & S^J_{\text{out},m,n}(\omega) = \frac{R_q^2}{4\pi} \int_{-\infty}^{\infty}d\omega_1 S_{\text{out},m, n}(\omega_1)  S_{\text{out},m, n}(\omega-\omega_1).
\end{align}
\end{subequations}
The cross-correlations~\eqref{eq:crossheat} are thus related to the charge-current cross-correlations~\eqref{eq:S_cross_explicit} and, hence, they do not depend on the chosen pair of channels $m,n$. Moreover, $S^J_{\text{out},m,n}(\omega)$ vanishes for global equilibrium, i.e., for $T_{\text{in}}=T_{\rm oc}=T$, since then $S_{\text{out},m, n}(\omega)=0$ according to Eq.~\eqref{eq:S_cross_explicit}. In what follows, we focus on the behaviour of the low-frequency components of the output heat-current noise, $\omega\ll T_\mathrm{in},T_\mathrm{oc},\tau_C^{-1}$. To this end, we consider the same two configurations as in previous sections.

\begin{figure}[t!]
\begin{center}
\captionsetup[subfigure]{position=top,justification=raggedright}
\subfloat[]{
\includegraphics[width =0.95\columnwidth]{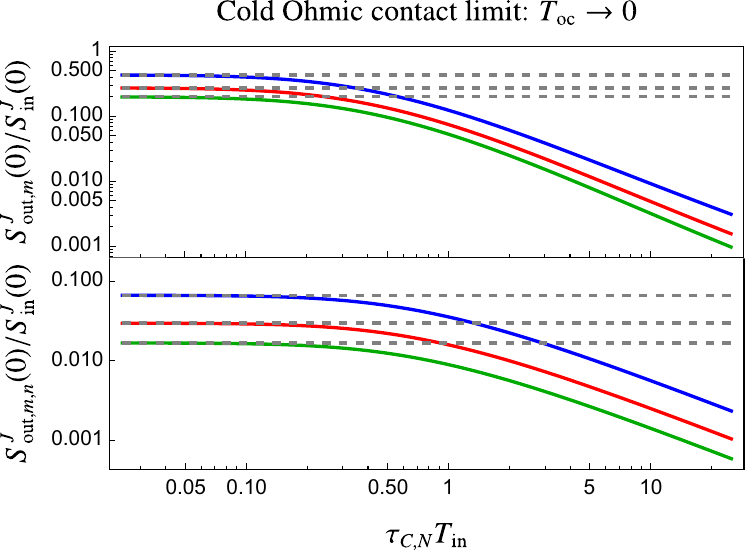}
\label{fig:heatnoise_comparisons_cold_contact_N}}
\\[-0.25cm]
\subfloat[]{
\includegraphics[width =0.95\columnwidth]{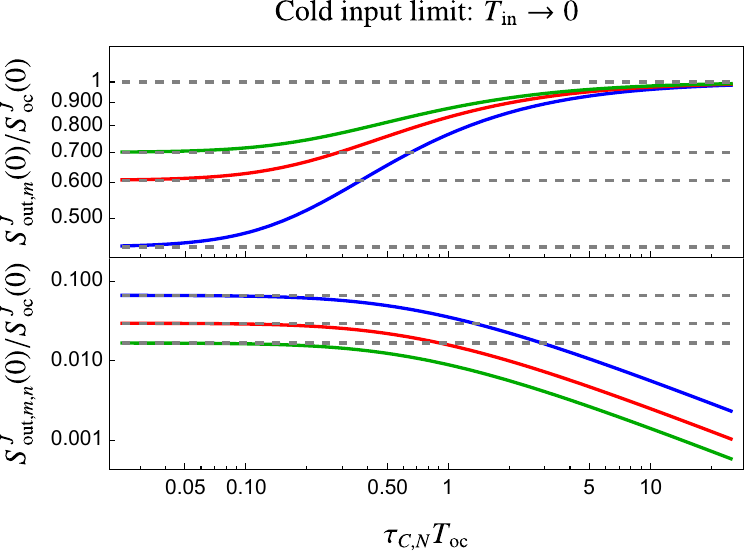}
\label{fig:heatnoise_comparisons_cold_input_N}}

\caption{ 
(a) Ratio of the low frequency heat current auto- and cross correlations~\eqref{eq:SQm_out_c1} and \eqref{eq:crossheat_c1} vs $\tau_{C,N}T_{\rm in}$ in the limit of a cold OC. Blue, red, and green curves depict cases with $N=2,3$, and $4$ pairs of attached channels, respectively. Dashed lines are the limits~\eqref{eq:heatnoiseratio1_N} and Eq.~\eqref{eq:heatnoiseratio2_N} for $\tau_{C,N}T_{\rm in}\ll 1$.   
(b) Ratio of the low frequency heat current auto- and cross correlations~\eqref{eq:SQm_out_c2} and \eqref{eq:crossheat_c2} vs $\tau_{C,N}T_{\rm oc}$ in the limit of cold input channels. Dashed lines are the limits~\eqref{eq:heatnoiseratio3_N} and \eqref{eq:heatnoiseratio4_N} for $\tau_{C,N}T_{\rm oc}\ll 1$ and unity for $\tau_{C,N}T_{\rm oc}\gg 1$. }
\label{fig:HeatNoise_comparisons_N}
\end{center}
\end{figure}

In configuration (i), we consider a very cold OC and take the limit $T_{\rm oc}\ll T_\mathrm{in}$ in the output heat-current auto- and cross-correlations~\eqref{eq:SQ_out_combined}.  After some algebraic manipulations, we obtain the integral expressions
\begin{subequations}
\label{eq:SJ_ColdOC_N}
\begin{align}
\label{eq:SQm_out_c1}
     &S_{\text{out},m}^{J}(0)= \frac{T_{\rm in}^3}{2\pi}\int_{0}^{\infty}dz 
     \frac{ z^2 \left(1+\left(e^z-1\right) g(z,T_{\rm in})\right)}{ \left(e^z-1\right)^2 \left(g(z,T_{\rm in})\right)^2},\\
\label{eq:crossheat_c1}
  & S^J_{\text{out},m,n}(0)= \frac{T_{\rm in}^3}{2\pi} \int_{0}^{\infty}d z 
   \frac{ z^2}{ \left(e^z-1\right)^2 \left(g(z,T_{\rm in})\right)^2},
\end{align}
\end{subequations} where we defined  $g(z,T) \equiv N \left(\left(\tau_{C,N} T\right)^2 z^2+1\right)$.
We plot the expressions~\eqref{eq:SJ_ColdOC_N} in Fig.~\ref{fig:heatnoise_comparisons_cold_contact_N}, 
and we see that the heat-current noises decay to zero for $\tau_{C,N}T_{\rm in}\gg 1$ but saturate to constant values for $\tau_{C,N}T_{\rm in}\ll 1$. We find these asymptotic values analytically as
\begin{align}
\label{eq:heatnoiseratio1_N}
    \frac{S_{\text{out},m}^{J}(0)}{S_{\text{in}}^{J}(0)} \approx
\begin{cases}
 \frac{\pi ^2+6 (N-1) \zeta (3)}{\pi ^2 N^2} & \tau_{C,N}T_{\rm in}\ll 1, \\ 
  \frac{3}{4   \pi  N^2} (\tau_{C,N} T_{\rm in})^{-1} &   \tau_{C,N}T_{\rm in} \gg 1
  \end{cases}
  \end{align}
and
\begin{align}
\label{eq:heatnoiseratio2_N}
    \frac{S_{\text{out},m,n}^{J}(0)}{S_{\text{in}}^{J}(0)} \approx
\begin{cases}
 \frac{\pi^2-6\zeta (3)}{\pi^2N^2} & \tau_{C,N}T_{\rm in}\ll 1, \\ 
  \frac{3}{4   \pi  N^2} (\tau_{C,N} T_{\rm in})^{-1} &   \tau_{C,N}T_{\rm in} \gg 1,
  \end{cases}
  \end{align}
with $S^J_{\rm in}(0)=\kappa_0 T_{\rm in}^3$ and $\zeta(3)\approx 1.2$. That both $S_{\text{out},m}^{J}(0)$ and $S_{\text{out},m,n}^{J}(0)$ become negligible with respect to $S_\mathrm{in}^J(0)$ for $\tau_{C,N}T_{\rm in} \gg 1$ indicates that the OC efficiently equilibrates the input channels so that output channels emanate close to the very cold OC distribution. In contrast, for $\tau_{C,N}T_{\rm in} \ll 1$, the OC's ability to equilibrate the input channels is suppressed, but not fully impeded. This can be seen from the fact that $S_{\mathrm{out},m}^J(0)$ is not negligible with respect to $S_\mathrm{in}^J(0)$, but nonetheless modified with respect to it, in contrast to the single channel case, where $S_{\mathrm{out},m}^J(0)\to S_\mathrm{in}^J(0)$. At the same time, non-vanishing cross-correlations are a manifestation of a nonequilibrium distribution in the outgoing channels, given by a mixture of input and OC distributions. 

For configuration (ii), we take instead $T_{\rm in}\ll T_\mathrm{oc}$ in Eq.~\eqref{eq:SQ_out_combined} and find the expressions
\begin{subequations}
\label{eq:SQm_out_cold_in}
    \begin{align}   
     &S_{\text{out},m}^{J}(0)\notag  \\ &= \frac{T_{\rm oc}^3}{2\pi}\int_{0}^{\infty}dz 
    \frac{ z^2 \left(1-g(z,T_{\rm oc}) \left(1+e^z\right)+\left(g(z,T_{\rm oc})\right)^2 e^z\right)}{ \left(e^z-1\right)^2 \left(g(z,T_{\rm oc})\right)^2},\label{eq:SQm_out_c2}\\ 
   & S^J_{\text{out},m,n}(0) = \frac{T_{\rm oc}^3}{2\pi} \int_{0}^{\infty}d z 
   \frac{ z^2}{ \left(e^z-1\right)^2 \left(g(z,T_{\rm oc})\right)^2},\label{eq:crossheat_c2}
 \end{align}
\end{subequations}
We note that the cross correlation~\eqref{eq:crossheat_c2} is equal to that in Eq.~\eqref{eq:crossheat_c1} upon substituting $T_{\rm oc}\leftrightarrow T_{\rm in}$. 
This happens because the two considered limits, $T_{\rm oc}\ll T_{\rm in}$ and $T_{\rm oc}\gg T_{\rm in}$, cause either $S_{\rm in}(\omega)$ or $S_{\rm oc}(\omega)$ to dominate in Eq.~\eqref{eq:S_cross_explicit}.
The resulting cross-correlations are thus determined by only one of these two noises and are otherwise identical except for an overall sign; they then enter quadratically in Eq.~\eqref{eq:crossheat}. 
We plot the expressions~\eqref{eq:SQm_out_cold_in} in
Fig.~\ref{fig:heatnoise_comparisons_cold_input_N}, and see that $S^J_{\text{out},m}(0)$ is strongly suppressed with respect to the OC noise for $\tau_{C,N}T_{\rm oc}\gg 1$ and that it saturates to a constant value for $\tau_{C,N}T_{\rm oc}\ll 1$. 
The cross-correlation noise $S^J_{\text{out},m,n}(0)$ is suppressed with respect to $S^J_{\rm oc}(0)$ for $\tau_{C,N}T_{\rm oc}\gg 1$ and approaches a constant value smaller than $S^J_{\rm oc}(0)$ for $\tau_{C,N}T_{\rm oc}\ll 1$. These limiting characteristics are given by the expressions
\begin{align}
\label{eq:heatnoiseratio3_N}
    \frac{S_{\text{out},m}^{J}(0)}{S_{\text{oc}}^{J}(0)} \approx
\begin{cases} 
  \frac{(N-1) \left((N-1) \pi ^2+6 \zeta (3)\right)}{N^2 \pi ^2} &   \tau_{C,N}T_{\rm oc} \ll 1, \\
  1+\frac{3 (1-4 N)(\tau_{C,N} T_{\rm oc})^{-1}}{4 \pi N^2} & \tau_{C,N}T_{\rm oc}\gg 1
  \end{cases}
  \end{align}
and
\begin{align}
\label{eq:heatnoiseratio4_N}
    \frac{S_{\text{out},m,n}^{J}(0)}{S_{\text{oc}}^{J}(0)} \approx
\begin{cases}
\frac{\pi^2-6\zeta (3)}{\pi^2N^2} & \tau_{C,N}T_{\rm oc}\ll 1, \\ 
  \frac{3}{4   \pi  N^2} (\tau_{C,N} T_{\rm oc})^{-1} &   \tau_{C,N}T_{\rm oc} \gg 1,
  \end{cases}
  \end{align}
with $S^J_{\rm oc}(0)=\kappa_0 T_{\rm oc}^3$.  Equations~\eqref{eq:heatnoiseratio3_N} and~\eqref{eq:heatnoiseratio4_N} show that in configuration (ii) the limit $\tau_{C,N}T_{\rm oc}\gg 1$ causes the OC to efficiently equilibrate the cold input channels and all output channels emanate with the hotter OC distribution. 
For $\tau_{C,N}T_{\rm oc}\ll 1$ the equilibration is instead suppressed, but not fully impeded. The non-vanishing cross-correlation further reflects the fact that the output channels are characterized by an out-of-equilibrium distribution, arising from a mixture of the input and OC distributions.

 The auto-correlation results presented in this subsection, for $N\to1$ all reduce to the single-channel limits given in Eqs.~\eqref{eq:SQout_integral1}-\eqref{eq:SQout_asymptotic_2}, as expected. 
Furthermore, we see from Eqs.~\eqref{eq:heatnoiseratio1_N},~\eqref{eq:heatnoiseratio2_N},~\eqref{eq:heatnoiseratio3_N}, and~\eqref{eq:heatnoiseratio4_N} that the presence of $N>1$ channels aids equilibration in comparison to the case of $N=1$, see Eq.~\eqref{eq:SQout_asymptotic_1} and~\eqref{eq:SQout_asymptotic_2}. This happens since the $N-1$ neutral modes of the input channels in the multi-channel setup always equilibrate efficiently. This feature was demonstrated in Ref.~\cite{slobodeniuk_equilibration_2013} for the charge-current noise. We have here explicitly demonstrated that this effect impacts also in the heat-current noise and how it manifests.  

In addition to the presented cross-correlations between output currents in different channels $m,n$, we find that---similarly to the single-channel case [cf. Eq.~\eqref{eq:SQio}]---there are non-vanishing cross-correlations between incoming and outgoing heat-current fluctuations of any channels $m,n$. 
To find them, we analyze the fast source term $\delta J_{\text{out},m}^{}(t)$ in Eq.~\eqref{eq:heatfluct_Ncase} and compute the correlation function
\begin{multline}
\label{eq:heat_cross_N}
  \left\langle \delta J_{\text{in},m} (\omega) \delta J_{\text{out},n}^{} (\omega')\right\rangle  \equiv 2\pi \delta(\omega+\omega') S_{\text{io},m,n}^{J}(\omega).
\end{multline}
As above, we choose all input temperatures to be equal, $T_{\text{in},m}=T_{\rm{in}}$, so that the heat-current cross-correlations $S_{\text{io},m,n}^{J}(\omega)$ do not depend on the input or output channel indices $m$ and $n$. The asymptotic values, analogous to~\eqref{eq:io_asym2}, of the zero-frequency heat-current cross-correlations~\eqref{eq:heat_cross_N} are then given as
\begin{align}\label{eq:io_asym_Ncase}
   \frac{S_{\text{io},m,n}^{J}(0)}{S^J_{\rm in}(0)}  \approx\begin{cases}
   \frac{1}{N^2} - \frac{4\pi^2}{5 N^2}(\tau_{C,N} T_{\rm in})^2 &    \tau_{C,N}  T_{\rm in} \ll 1,\\
\frac{3}{2\pi N^2} \frac{1}{\tau_{C,N} T_{\rm in}} & \tau_{C,N} T_{\rm in} \gg 1,
\end{cases}
\end{align}
with $S^J_{\rm in}(0)=\kappa_0 T_{\rm in}^3$. This result shows how the finite correlations between input and output heat current fluctuations are affected by the potential fluctuations on the time scale $\tau_{C,N}$ and, consequently, how the OC in the heat Coulomb-blockade regime acts as a source of heat current fluctuations in the absence of temperature fluctuations. 

\subsubsection{Impact of potential fluctuations on the total heat conductance: multi-channel heat Coulomb blockade}
\label{eq:N_channel_HCB}
Based on Eq.~\eqref{eq:Charge_dyn_N}, we establish in this subsection the heat Coulomb blockade effect in the heat conductance of the $N$-channel setup, confirming previous results in Refs.~\cite{Sivre2018Feb,sivre2019electrical}. Note that we here still consider the limit $\tau_E \rightarrow \infty$; sizeable temperature fluctuations are instead analyzed below in Sec.~\ref{sec:N_HeatDynamics}. 

\begin{figure}[t!]
\begin{center}
\captionsetup[subfigure]{position=top,justification=raggedright}
\subfloat[]{
\includegraphics[width=0.95\columnwidth]{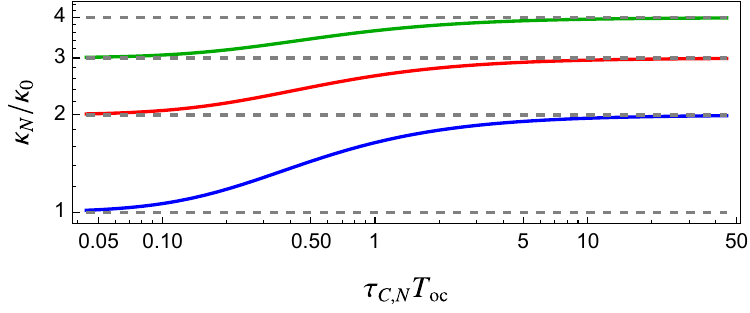}
\label{fig:kappa_renorm_multi}}
\\[-0.25cm]
\subfloat[]{
\includegraphics[width =0.95\columnwidth]{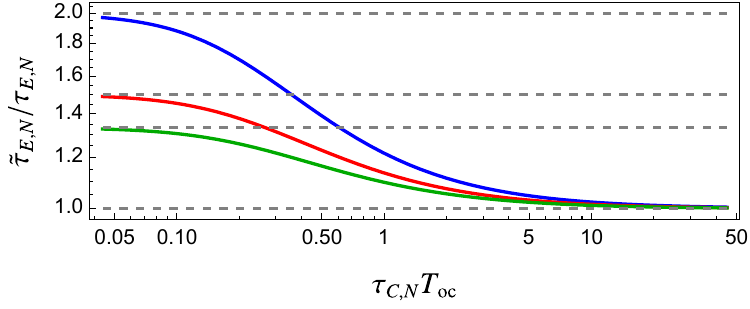}
\label{fig:tauE_renorm_multi}}
\caption{ (a)  Renormalization of the total $N$-channel heat conductance $\kappa_{N}T_{\rm oc}/(\kappa_0 T_{\rm oc})$ in Eq.~\eqref{eq:J_out_m} vs $\tau_{C,N} T_{\rm oc}$, where $\tau_{C,N}$ is the multi-channel OC $RC$ time and $T_{\rm oc}$ is the OC temperature. Blue, red, and green, depict the heat conductance when $N=2,3$, or $4$ pairs of channels, respectively, are connected.  Black and gray dashed lines depict the asymptotic limits $N-1$ and $N$. (b) Renormalization of the $N$-channel energy relaxation time $\tilde{\tau}_{E,N}/\tau_{E,N}$ in Eq.~\eqref{eq:tE_tilde_N}, vs $\tau_{C,N}T_{\rm oc}$. Blue, red, and green solid lines depict the energy relaxation time when $N=2,3$, or $4$ pairs of channels, respectively, are connected to the OC. Gray dashed lines depict the asymptotic limits $N/(N-1)$ for $\tau_{C,N}T_{\rm oc}\ll 1$ and the asymptotic limit of unity (independently of $N$) for $\tau_{C,N}T_{\rm oc}\gg 1$.}
\label{fig:multi_renorm}
\end{center}
\end{figure}

In the $N$-channel setup, the heat Coulomb blockade effect amounts to the suppression of exactly one measured heat conductance quantum~\cite{Sivre2018Feb,sivre2019electrical}.
To demonstrate this effect with our model, we compute the linear response coefficient of the $m$-th channel output heat current, $J_{\text{out},m}$,  when $N$ pairs of channels are attached to the OC. It is given as 
\begin{align}
\label{eq:J_out_m}
\kappa_m T_{\rm oc} & \equiv \frac{ \partial J_{\text{out},m} }{\partial T_{\rm oc} } =\kappa_0 T_{\rm oc}+\frac{1}{4 \pi N T_{\rm oc} \tau_{C,N}^2}\notag \\ &+\frac{1}{4 N \tau_{C,N}}-\frac{\psi'\left(\frac{1}{2 \pi  T_{\rm oc} \tau_{C,N}}\right)}{8 \pi^2 N T_{\rm oc}^2 \tau_{C,N}^3}\ ,
\end{align}
where we inserted the finite-frequency noise~\eqref{eq:Sout_Ncase_2} into the heat-current expression~\eqref{eq:T_integral} and then differentiated with respect to $T_{\rm oc}$. In order to get insights into the behavior of the heat conductance, we consider the following two limiting cases of Eq.~\eqref{eq:J_out_m},
\begin{align} \label{eq:thermal_cond_Ncase}
    \frac{\kappa_m}{\kappa_0}
    = \begin{cases}
        1-\frac{1}{N}+\frac{4 \pi ^2}{5 N}   \tau_{C,N}^2 T_{\rm oc}^2& \tau_{C,N} T_{\rm oc} \ll 1,\\
        1-\frac{3}{2\pi} \frac{1}{N \tau_{C,N} T_{\rm oc} } & \tau_{C,N} T_{\rm oc} \gg 1.
    \end{cases} 
\end{align} 
Summing up the equal contributions from each of the $N$ output channels, we find that the total heat conductance in the limit $\tau_{C,N} T_{\rm oc}\ll 1$ becomes
\begin{align}
\label{eq:kappa_tot_reduced}
    \kappa_{N}T_{\rm oc} \equiv N\kappa_m T_{\rm oc}=\kappa_0(N-1)T_{\rm oc}.
\end{align}
Hence, the total heat conductance is reduced by \textit{precisely one unit}. 
This reduction can be interpreted as the \textit{total charge mode} (with resistance $R_q/N$) of the impinging channels being teleported across the OC~\cite{Sivre2018Feb,Duprez2019} thereby retaining its incoming distribution function.
Hence, the heat current of this single effective mode has no $T_{\rm oc}$-dependence and therefore drops out in the heat conductance. 
By contrast, the $N-1$ impinging \textit{neutral} (or ``dipole'') modes emanate with the OC distribution function and thus contribute with $N-1$ units to the net heat conductance. 
For $N=1$,  the heat conductance vanishes, in agreement with the single channel case~\eqref{eq:thermal_cond}, where the single channel coincides with the charge mode. In the opposite limit, $\tau_{C,N} T_{\rm oc}\gg 1$, we find instead
\begin{align}
\label{eq:ktot_other_limit}
    \kappa_{N}T_{\rm oc} = N\kappa_mT_{\rm oc} = \left(N\kappa_0 - \frac{3\kappa_0}{2\pi\tau_{C,N}T_{\rm oc}}\right)T_{\rm oc}
\end{align}
i.e., there is a small correction per channel which decreases with increasing $\tau_{C,N}T_{\rm oc}$, with $\kappa_N$ eventually reaching $\kappa_{N}=N\kappa_0$.  The correction term can be written in terms of the OC single particle charging energy $E_C=\pi \tau_C^{-1}$ as
\begin{align}
    \frac{3\kappa_0}{2\pi\tau_{C,N}T_{\rm oc}} = \frac{3\kappa_0}{2\pi^2T_{\rm oc}}\times NE_C.
\end{align}
which thus increases linearly with increasing charging energy and the number of attached channels $N$~\cite{sivre2019electrical}. We plot the heat conductance ratio $\kappa_{N}/\kappa_0\equiv N\kappa_m/\kappa_0$, with $\kappa_m$ from Eq.~\eqref{eq:J_out_m}, in Fig.~\ref{fig:kappa_renorm_multi}, explicitly displaying the full crossover from $N$ to $N-1$ channels contributing to the total heat conductance. 

\subsection{Multi-channel heat dynamics}
\label{sec:N_HeatDynamics}
We now release the constraint of an infinitely large energy relaxation time $\tau_E$ and investigate the role of temperature fluctuations on the heat dynamics of the $N$-terminal setup for $\tau_E\gg\tau_C$, but still finite.

\subsubsection{Multi-channel Langevin equations for heat-current fluctations and OC temperature fluctuations}
Having computed the correlation function~\eqref{eq:Sio_def} of the fast heat-current fluctuations $\delta J_\mathrm{out}$ as well as the heat conductance~\eqref{eq:J_out_m}, we now solve Eqs.~\eqref{eq:NChannel_heat_Dynamics} for the total output heat-current fluctuations. The solution reads
\begin{subequations}
\begin{align}
    &\delta J^{\rm tot}_{\text{out},m}(\omega) = \sum_{p=\rm{in},\rm{out}}\sum_{n=1}^N\mathcal{T}^J_{p,mn}(\omega)\delta J_{p,n}(\omega),\label{eq:deltaJSol_Ncase}\\
    & \mathcal{T}^J_{\text{out},mn}(\omega) \equiv  \delta_{mn}  - N^{-1}\left[1-i\omega  \tilde{\tau}_{E,N} \right]^{-1}
    \\
    & \mathcal{T}^J_{\text{in},mn}(\omega) \equiv  N^{-1}\left[1-i\omega  \tilde{\tau}_{E,N} \right]^{-1}.
    \label{eq:tq_values_Ncase}
\end{align}
\end{subequations}
Using these expressions, we obtain the sought-for total heat-current noise $S_{\text{out},m}^{J,\text{tot}}(\omega)$, 
\begin{subequations}
\begin{align}
    & \langle \langle \delta J^{\rm tot}_{\text{out},m}(\omega) \delta J^{\rm tot}_{\text{out},m}(\omega')\rangle \rangle_E \equiv 2\pi \delta (\omega+\omega') S_{\text{out},m}^{J,\text{tot}}(\omega)
\end{align}
which provides the multi-channel generalization of Eq.~\eqref{eq:SJ_tot_single}, namely
\begin{align}
    &S_{\text{out},m}^{J,\text{tot}}(\omega) = \notag \\ & \frac{S^J_{\rm in}(0)+S^J_{\text{out},m}(0)\left[N-1+N\omega^2\tilde{\tau}_{E,N}^2\right]}{N(1+\omega^2\tilde{\tau}_{E,N}^2)}
    - \frac{S^J_{\text{out},m,n}(0)}{1+\omega^2 \tilde{\tau}_{E,N}^2}\ .\label{eq:SoutQ_NCase}
\end{align}
\end{subequations}
The characteristic energy relaxation time entering in Eq.~\eqref{eq:SoutQ_NCase} is the total energy relaxation time of the OC, namely
\begin{align}
\label{eq:tE_tilde_N}
    \tilde{\tau}_{E,N} \equiv \frac{\tilde{\tau}_{E,m}}{N},
\end{align}
where $\tilde{\tau}_{E,m} = C_E[T_{\rm oc}]/(\kappa_m T_{\rm oc})$ for the multi-channel case with $\kappa_m$ from Eq.~\eqref{eq:thermal_cond_Ncase}. We will discuss more details concerning $\tilde{\tau}_{E,m}$ in Sec.~\ref{sec:Non_Gaussian} below. In deriving Eq.~\eqref{eq:SoutQ_NCase}, we used that when the temperature fluctuations are taken into account, the contributions from the input and OC heat-current noises in Eq.~\eqref{eq:SoutQ_NCase} are given by $S^J_p(0)$, for $p\in\{\rm{in},\rm{out}\}$, paralleling the discussion on the time scale separation in Sec.~\ref{sec:HeatNoise_Temp_Fluct}. 

We see that depending on $\omega \tilde{\tau}_{E,N}$, the total output heat noise in Eq.~\eqref{eq:SoutQ_NCase} interpolates between 
\begin{align}\label{eq:SJ_tot_interpolation}
   & S_{\text{out},m}^{J,\text{tot}}(\omega) \notag \\
   &=\begin{cases}
   \frac{S^J_{\rm in}(0)+S^J_{\text{out},m}(0)\left[N-1\right]- NS^J_{\text{out},m,n}(0)}{N}
     &    \omega\tilde{\tau}_{E,N}\ll 1,\\
S^J_{\text{out},m}(0)
     &  1 \ll\omega\tilde{\tau}_{E,N}, 
\end{cases}
\end{align}
where we always have $\omega\tau_{C,N}\ll1,\omega\tilde{\tau}_{E,N}$. This total heat current noise depends on the fluctuations of the OC Langevin source $S_\mathrm{oc}(0)$ through the outgoing heat-current noise $S^J_{\mathrm{out},m}$. 
In the limit $\omega\tilde{\tau}_{E,N}\gg 1$, the output heat noise is not impacted by temperature fluctuations, but only by voltage fluctuations, as discussed for the single-channel case, Eq.~\eqref{eq:SJ_tot_single}, in Sec.~\ref{sec:heatnoise_voltage_N}. For frequencies smaller than the inverse modified energy relaxation time, temperature fluctuations did completely mask the effect of $S_\mathrm{oc}^J(0)$ in the single-channel case; this is very different in the multi-terminal case, where both $S^J_{\text{out},m}(0)$ and $S^J_{\text{out},m,n}(0)$, and hence also $S_\mathrm{oc}^J(0)$ contribute to the total heat-current noise even when temperature fluctuations are sizable. Formally, this is again similar to the multi-terminal heat Coulomb blockade induced by voltage fluctuations, however, being due to the \textit{heat} capacitance, it has a physically very different origin here.

The sum rule~\eqref{eq:heat_sumrule} holds also in the $N$-channel setup and reflects energy conservation.
By using this sum rule, we see that the output heat-current noise~\eqref{eq:SoutQ_NCase} has equilibrium form for $T_{\rm in}= T_{\rm oc}=T$. The heat-current cross-correlations~\eqref{eq:crossheat} that enter in the final term in  Eq.~\eqref{eq:SoutQ_NCase} are given by the convolution of the charge-current cross-correlations~\eqref{eq:S_cross_def} with themselves. This contribution thus vanishes in global equilibrium, as explicitly seen from the definition~\eqref{eq:S_cross_explicit}.

We also compute the  heat-current cross-correlations, defined for $m\neq n$ as
\begin{align}
\label{eq:SQ_cross_def}
    \langle\langle \delta J^{\rm tot}_{\text{out},m}(\omega) \delta J^{\rm tot}_{\text{out},n}(\omega') \rangle \rangle_E \equiv 2\pi \delta(\omega+\omega')S^{J,\text{tot}}_{\text{out},m,n}(\omega),
\end{align}
and obtain
\begin{align}
\label{eq:SQ_cross_explicit}
    S^{J,\text{tot}}_{\text{out},m,n} (\omega) = \frac{S^J_{\rm in}(0)-S^J_{\text{out},m}(0)}{N(1+\omega^2\tilde{\tau}_{E,N}^2)}+ \frac{S^J_{\text{out},m,n}(0) \omega^2 \tilde{\tau}_{E,N}^2}{1+\omega^2\tilde{\tau}_{E,N}^2}.
\end{align}
We find a relation similar to that between charge-current auto- and cross-correlations, Eq.~\eqref{eq:relation_corr}, also for the heat-current fluctuations due to the structural similarity of the equations. This relation reads
\begin{align}\label{eq:relation_corr_heat}
    S^{J,\text{tot}}_{\text{out},m,n} (\omega)
    = S^{J,\text{tot}}_{{\rm out},m}(\omega) +S^J_{\text{out},m,n}(0)-S^J_{\text{out},m}(0).
\end{align}
The interpolating behaviour of the cross correlations is
\begin{align}\label{eq:SJ_tot_interpolation2}
   & S_{\text{out},m,n}^{J,\text{tot}}(\omega) =\begin{cases}
   \frac{S^J_{\rm in}(0)-S^J_{\text{out},m}(0)}{N}
     &    \omega\tilde{\tau}_{E,N}\ll 1,\\
S^J_{\text{out},m,n}(0)
     & \omega\tilde{\tau}_{E,N}\gg 1.
\end{cases}
\end{align}
Here, the former limit shows how the OC temperature fluctuations impact the heat-current cross-correlations, while the latter limit manifests the situation in Sec.~\ref{sec:heatnoise_voltage_N} where temperature fluctuations are fully neglected.

\subsubsection{Impact of heat Coulomb blockade on the energy relaxation time}\label{sec:N-terminal_tau_E}
Here, we analyze the characteristics of the modified energy relaxation time, focussing on the heat Coulomb blockade regime, i.e., with the heat conductance~\eqref{eq:thermal_cond_Ncase} taken in the limit $\tau_{C,N} T_{\rm oc} \ll 1$. Then, to leading order in $\tau_{C,N} T_{\rm oc}$, we find that the energy relaxation time~\eqref{eq:tE_tilde_N} becomes
\begin{align} 
\label{eq:tau_E_blocked}
    \tilde{\tau}_{E,N} \approx \frac{C_E[T_{\rm oc}]}{(N-1)\kappa_0 T_{\rm oc}}, \quad N > 1.
\end{align}
We thus see that the contribution from exactly one channel is missing due to the reduced heat conductance~\eqref{eq:kappa_tot_reduced}. This enhancement of the energy relaxation time $\tilde{\tau}_{E,N}/\tau_{E,N}$ is plotted in Fig.~\ref{fig:tauE_renorm_multi}.  
In contrast to the single-channel case~\eqref{eq:New_tau_E}, the ratio $\tilde{\tau}_{E,N}/\tau_{E,N}$ does not diverge in the limit $\tau_{C,N} T_{\rm oc}\rightarrow 0$, but saturates instead to the value $\tilde{\tau}_{E,N}/\tau_{E,N}=N/(N-1)$, see Fig.~\ref{fig:tauE_renorm_multi}.
As described above in Sec.~\ref{eq:N_channel_HCB}, it is only the total charge mode (i.e., the charged sector of the edge channels) that is affected by the heat Coulomb blockade. 
We see here that the blockade implies that this mode does not contribute to the enhancement of the energy relaxation time either, as it  ``teleports''~\cite{Idrisov2018Jul,Duprez2019} through the OC and does therefore not aid in relaxing the OC energy. In contrast, the $N-1$ neutral modes efficiently contribute to the OC energy relaxation. 

\subsubsection{OC temperature fluctuations in the multi-channel setup}
The modified energy relaxation time, naturally, has a direct impact on the OC temperature fluctuations. To show it's impact, we compute the OC temperature-fluctuations correlation function by using the solution of the Langevin equation~\eqref{eq:deltaJSol_Ncase} in combination with Eq.~\eqref{eq:heatfluct_Ncase}, and find
\begin{align}
    &\langle\left\langle \Delta T_{\rm oc}(\omega) \Delta T_{\rm oc}(\omega')\right\rangle\rangle_E \equiv 2\pi \delta(\omega+\omega')S^T_{\rm oc}(\omega),\label{eq:TT_corr_multi} \\ 
    &S^T_{\rm oc}(\omega) = \frac{N^2}{\kappa_N^2 T_{\rm oc}^2} \notag \\
    &\times \frac{S^J_{\rm in}(0) - 2 N S^J_{\text{io},m,n}(0) + N S^J_{\text{out},m,n}(0) + S^J_{\text{out},m}(0)}{N(1+\omega^2 \tilde{\tau}_{E,N}^2)}.\label{eq:TT_corr_multi_1b}
\end{align}
Similarly to the single-channel case, we consider for simplicity only the case of equal temperatures $T_{\rm in} = T_{\rm oc}=T$. In this case, heat-current cross-correlations $S^J_{\text{out},m,n}(0)$ vanish from~\eqref{eq:TT_corr_multi_1b}. Next, by using the asymptotic form of $S^J_{\text{io},m,n}(0)$, given in~\eqref{eq:io_asym_Ncase}, we obtain
\begin{align}
\label{eq:TT_corr_multi_2}
    S^T_{\rm oc}(\omega)=\frac{1}{\kappa_N} \frac{2T}{(1+\omega^2 \tilde{\tau}_{E,N}^2)},
\end{align}
 due to a cancellation similar to that for the single channel case in  Eq.~\eqref{eq:temp_identity}. Explicitly, we have
\begin{multline}
\label{eq:temp_identity_NCase}
    S_{\rm in}^J(0) - 2 N S_{\text{io},m,n}^J(0)+S_{\text{out},m}^J(0) \\= R_q^2 \int_{-\infty}^{\infty}\frac{d\omega'}{2\pi} \left(1-N\left|\mathcal{T}_{\text{in},mn}(\omega)\right|^2\right) S(\omega') S(-\omega') \\=  T^2 \partial_T R_q \int_{-\infty}^{\infty}\frac{d\omega'}{2\pi} \left|\mathcal{T}_{\text{oc},mn}(\omega)\right|^2  S(\omega) = 2 \kappa_m T^3,
\end{multline}
independently of $\tau_{C,N} T$. In the time domain, Eq.~\eqref{eq:TT_corr_multi_2} becomes
\begin{align} \label{eq:temp_supp_N}
    \left\langle \left\langle 
    \Delta T_{\rm oc} (t) \Delta T_{\rm oc} (0) \right\rangle \right\rangle_E  &= \left\langle \left\langle 
    \Delta T(t) \Delta T(0) \right\rangle \right\rangle_E \notag \\   &=\frac{T^2}{C_E[T]} e^{-|t|/\tilde{\tau}_{E,N}}.
\end{align} 
Hence, similarly to the single channel case~\eqref{eq:temp_supp}, the amplitude of the OC temperature-fluctuations correlation function is universal~\cite{Pekola21Oct}, but its decay rate, here $\tilde{\tau}_{E,N}$, depends even in equilibrium on the charge relaxation time via Eq.~\eqref{eq:thermal_cond_Ncase} and Eq.~\eqref{eq:tE_tilde_N}. 

\section{Non-Gaussian full counting statistics due to temperature fluctuations}
\label{sec:Non_Gaussian}
When the OC temperature fluctuations are negligible, it has been shown, see e.g., Ref.~\cite{slobodeniuk_equilibration_2013}, that the edge channel-OC system is well described by a quadratic bosonized model, as briefly reviewed in Appendix~\ref{sec:Appendix_B}. This quadratic nature implies that the full counting statistics (FCS)~\cite{Levitov1993Aug,Levitov1996Oct}
 of the output charge and heat currents only contain first and second cumulants, i.e., the FCS is Gaussian. This can feature can change when temperature fluctuations come into play. Indeed, it has been shown~\cite{vandenBerg2015Jul} that the FCS of the emitted \textit{energy} show non-Gaussian correlations in the presence of temperature fluctuations.
 In this section, we demonstrate a novel and complementary effect of sizeable OC temperature fluctuations, namely a non-Gaussian FCS of the output \textit{charge} current, induced by temperature fluctuations.

In order to calculate cumulants of the FCS of the charge current, we focus on the charge current in one of the output channels, $m$, in the $N$-channel setup, see Fig.~\ref{fig:Multi_Channel_Setup}. We define the net electrical charge fluctuations $\delta q$, that traverse the cross-section of this channel in the observation time interval $\tau_{\rm obs}$ as
\begin{align}
\label{eq:charge_obs}
      \delta q(\tau_{\rm obs}) \equiv \int\limits_{-\tau_{\rm obs}/2}^{\tau_{\rm obs}/2} \hspace{-.35cm} dt \ \delta I_{\text{out},m}(t)\equiv\int\limits_{-\tau_{\rm obs}/2}^{\tau_{\rm obs}/2} \hspace{-.35cm} dt \ \delta I(t).
\end{align}
Here, we assume the time interval of observation to be very long, meaning that we focus on the zero-frequency FCS. For notational ease, we denote $\delta I_{\text{out},m}(t)\equiv\delta I(t)$ in  Eq.~(\ref{eq:charge_obs}) and in the remainder of this section. 

Without temperature fluctuations, the system is, as mentioned above, Gaussian, and statistical quantities, like the third and fourth cumulant of the transferred charge fluctuations
\begin{subequations}
\label{eq:cumulants}
\begin{align}
\label{eq:thid_cumulant}
& \mathcal{C}_3 \equiv \left\langle (\delta q(\tau_{\rm obs}))^3 \right \rangle,\\
\label{eq:fourth_cumulant}
& \mathcal{C}_4 \equiv \left\langle (\delta q(\tau_{\rm obs}))^4  \right \rangle -3 \left( \left\langle (\delta q(\tau_{\rm obs}))^2  \right \rangle \right)^2,
\end{align}
\end{subequations}
vanish due to Wick's theorem. However, in the presence of OC temperature fluctuations, we show in the remainder of this section that the different time scales of charge end energy dynamics produce additional non-zero contributions, which we denote as $\Delta \mathcal{C}_j$, for $j=3,4$, to the right-hand sides of Eq.~\eqref{eq:cumulants}.

These contributions can be thought of as ``noise of noise'' contributions and are very similar to so-called ``cascade corrections''  considered previously in Refs.~\cite{Pilgram_2003,Nagaev2002Nov,Pilgram2004Jul,vandenBerg2015Jul}. 
These earlier works studied the FCS of a chaotic cavity, which is a large conductor connected to two metallic leads with their number of transmission channels being controllable by quantum point contacts. Analogously to the edge channel-OC system considered here, such a cavity is characterized by both an $RC$ time and by slow fluctuations of its particle distribution function (in the present work, these slow fluctuations are instead produced by temperature fluctuations). The results we obtain in this section are  comparable to the inelastic ``hot-electron regime" in the previously studied case of a cavity with symmetrically connected channels. However, in contrast to previous works, we consider here only fully chiral edge channels as the connected transmission channels. Still, the case of $N=2$ chiral channels is very closely related to two non-chiral channels symmetrically connected to the cavity, and our results in this limit agree with the ones found in Ref.~\cite{Nagaev2002Nov}. 

\subsection{Fourth cumulant}
In order to compute the correction, $\Delta \mathcal{C}_4$, to $\mathcal{C}_4$ in Eq.~\eqref{eq:fourth_cumulant} we explicitly include the time-dependent fluctuations of the OC temperature, $T_{\rm oc}(t)=T_{\rm oc}+\Delta T_{\rm oc}(t)$, in the two-point charge-current correlation functions. We therefore write the correction to the fourth cumulant as the leading order term in a functional derivative expansion, using the double averaging procedure introduced in Sec.~\ref{sec:BL_approach}. We find that the correction is given by
\begin{align}
\label{eq:4C_corr}
     &\Delta \mathcal{C}_4 \equiv \prod_{n=1}^4 \int\limits_{-\tau_{\rm obs}/2}^{\tau_{\rm obs}/2} dt_n \Big(
    \left\langle\left\langle\delta I(t_1)\delta I(t_2)\delta I(t_3)\delta I(t_4)\right\rangle \right\rangle_E \notag \\ &-3  \left\langle\left\langle\delta I(t_1)\delta I(t_2)\right \rangle \right\rangle_E \left\langle \left \langle \delta I(t_3)\delta I(t_4)\right\rangle \right\rangle_E \Big) \notag \\
    &\approx 3\prod_{n=1}^2 \int\limits_{-\tau_{\rm obs}/2}^{\tau_{\rm obs}/2} dt_n     \left(\frac{\partial S_{\text{out},m}(0,T_{\rm oc})}{\partial T_{\rm oc}}\right)^2 \notag \\ &\left\langle\langle \Delta T_{\rm oc}(t_1)\Delta T_{\rm oc}(t_2) \right\rangle\rangle_E. 
\end{align}
Here, the factor of $3$ is the number of possible ways to pair the currents into two-point charge-current correlation functions. We also have that $S_{\text{out},m}(0,T_{\rm oc})$ (with the $T_{\rm oc}$ dependence explicitly highlighted here) is the low frequency output charge-current noise given in Eq.~\eqref{eq:Sout_Ncase_3}. In the following, we show how to derive the second step in Eq.~\eqref{eq:4C_corr}.

Equation~\eqref{eq:4C_corr} thus represents the non-vanishing leading order contribution to the fourth cumulant and arises due to a competition of dynamics on two different time scales (taken care of by the double averaging procedure). On time scales corresponding to an averaging over charge fluctuations, the charge current fluctuations are Gaussian. We may thus use Wick's theorem for these fluctuations so that 
\begin{align}
\label{eq:split_simplifcation}
     & \prod_{n=1}^4 \int\limits_{-\tau_{\rm obs}/2}^{\tau_{\rm obs}/2} dt_n \
     \left\langle\left\langle\delta I(t_1)\delta I(t_2)\delta I(t_3)\delta I(t_4)\right\rangle \right\rangle_E  \notag \\ &=  \prod_{n=1}^4 \int\limits_{-\tau_{\rm obs}/2}^{\tau_{\rm obs}/2} dt_n \
     \left\langle \left\langle \delta I(t_1)\delta I(t_2) \right\rangle  \left\langle\delta I(t_3)\delta I(t_4)\right\rangle \right\rangle_E.
\end{align}
Note that in the second line, the average $\langle...\rangle_E$ is taken over a product over two averages $\langle ... \rangle$. In the absence of temperature fluctuations, the correlation function $ \left\langle \delta I(t_1)\delta I(t_2) \right\rangle \equiv S(t_1,t_2)= S(t_1-t_2)$ is the short-hand notation for the Fourier transformed  output charge noise power~\eqref{eq:Sout_Ncase_2}, i.e, the output charge-current noise in the time-domain. We see that the time difference $t_1-t_2$ governs the dynamics of this quantity. However, due to the OC temperature fluctuations, the correlation function  $S(t_1,t_2)\neq S(t_1-t_2)$, i.e., it loses this time-translational invariance.

To account for this time-dependence, we now change the time integration variables in Eq.~\eqref{eq:split_simplifcation}  into the sum, $t_s =(t_1+t_2)/2$, and difference, $t_d =t_1-t_2$, of times (and similarly for $t_3$ and $t_4$). Even though not time-translation invariant anymore, the charge-current noise $S(t_1,t_2)$ is still peaked around zero with respect to the time difference $t_d$ on the time scale $\tau_{C,N}$. This makes us identify $t_d$ with the characteristic time scale of the charge dynamics. By contrast, the time dependence of the much slower OC temperature fluctuations must occur on the $t_d$-independent time scale $t_s$. For the first set of time integrals in Eq.~\eqref{eq:split_simplifcation}, we may thus write
\begin{align}
\label{eq:time_trafo}
     & \prod_{n=1}^2 \int\limits_{-\tau_{\rm obs}/2}^{\tau_{\rm obs}/2} dt_n \
     \left\langle S(t_1,t_2) \right\rangle_E  \notag \\&\approx   \int\limits_{-\tau_{\rm obs}}^{\tau_{\rm obs}} dt_{d} \int\limits_{-\tau_{\rm obs}/2+ |t_{d}|/2}^{\tau_{\rm obs}/2- |t_{d}|/2} dt_{s}  \  \left\langle S(t_d,T_{\rm oc}(t_s)) \right\rangle_E,
\end{align}
and similarly for the analogous integrals involving $t_3$ and $t_4$. 
Since the observation time $\tau_{\rm obs} \rightarrow \infty$ is the largest time scale, we next assume that the integration boundaries over $t_s$  are independent of $t_d$. 
This assumption can again be justified by the fact that we consider measurement frequencies $\omega \tau_{C,N} \ll 1$, such that $S(t_d,T_{\rm oc}(t_s))$ will be peaked around $t_d=0$ with a width given by $\tau_{C,N}$. On the short time-interval given by $\tau_{C,N}$, the temperature can further be assumed to be constant i.e., $T_{\rm oc}(t_s)\approx T_{\rm oc}$.
This means we can evaluate  the integral in Eq.~\eqref{eq:time_trafo} over $t_d$ only. The combination of the above approximations allows us simplify the following integral 
\begin{align}
\label{eq:2p_approx}
    &S_{\text{out},m}(0,T_{\rm oc}(t_s)) = \int\limits_{-\tau_{\rm obs}}^{\tau_{\rm obs}} dt_d\ S(t_d,T_{\rm oc}(t_s)) \notag \\ 
   &\approx  S_{\text{out},m}(0,T_{\rm oc}) + \int\limits_{-\infty}^\infty dt \frac{\delta S_{\text{out},m}(0,T_{\rm oc}(t_s))}{\delta T_{\rm oc}(t)} \Delta T_{\rm oc}(t) \notag\\&= S_{\text{out},m}(0,T_{\rm oc}) +  \frac{\partial S_{\text{out},m}(0,T_{\rm oc})}{\partial T_{\rm oc}} \Delta T_{\rm oc}(t_s)  
\end{align} 
Here, in the first line, we have integrated to obtain the zero-frequency noise. In the second and third lines, we expanded around the weak time-dependence of the OC temperature fluctuations. We now use the approximation~\eqref{eq:2p_approx} in Eq.~\eqref{eq:time_trafo} and insert it into Eqs.~\eqref{eq:split_simplifcation} and~\eqref{eq:4C_corr}. Then, inserting also the temperature-fluctuations correlation function in Eqs.~\eqref{eq:TT_corr_multi} and~\eqref{eq:TT_corr_multi_1b}, we find that the fourth cumulant~\eqref{eq:fourth_cumulant} evaluates to
\begin{multline}
\label{eq:finite_FC}
    \mathcal{C}_4 \approx \Delta \mathcal{C}_4 = 3 \tau_{\rm obs}\left(\frac{N-1}{R_q N}\right)^2  \frac{1- e^{-\tau_{\rm obs}/(2 \tilde{\tau}_{E,N})}}{N \kappa_m^2T_{\rm oc}^2 }
    \\\times\left(S^J_{\rm in}(0) - 2 N S^J_{\text{io},m,n}(0) + N S^J_{\text{out},m,n}(0) + S^J_{\text{out},m}(0)\right),
\end{multline} 
where the heat current noises on the right-hand side are given by Eq.~\eqref{eq:SQ_out_combined}, Eq.~\eqref{eq:heat_cross_N}  and $S^J_{\rm in}(0)=\kappa_0 T_{\rm in}^3$.
For global equilibrium conditions, $T_{\rm in} =T_{\rm oc} =T$, we may further use the heat-current noise identity~\eqref{eq:temp_identity_NCase} with which Eq.~\eqref{eq:finite_FC} reduces to
\begin{multline}
\label{eq:finite_FC_2}
    \mathcal{C}_4 \approx \Delta \mathcal{C}_4 = \frac{6 T \tau_{\rm obs}}{N\kappa_m}\left(\frac{N-1}{R_q N}\right)^2  \left(1- e^{-\tau_{\rm obs}/(2 \tilde{\tau}_{E,N})}\right).
\end{multline} 
We note that the fourth-cumulant correction terms~\eqref{eq:finite_FC} and~\eqref{eq:finite_FC_2} are both non-zero whenever $N\neq 1$ and $\tilde{\tau}_{E,N}$ [see Eq.~\eqref{eq:tE_tilde_N}] is finite. For the special case $N=1$, the OC temperature fluctuations do not produce a non-zero fourth cumulant, even for finite $\tilde{\tau}_{E,N}$. The reason is that for a single connected channel, the output charge current and zero-frequency charge current noise do not depend on $T_{\rm oc}(t)$ due to charge conservation, see Eqs.~\eqref{eq:I_out_I_in} and~\eqref{eq:ZeroFreqTemp}. The fourth cumulant is therefore, in this case, unaffected by the OC temperature fluctuations and vanishes.

With the fourt cumulant presented in Eq.~\eqref{eq:finite_FC}, we have thus demonstrated that finite temperature fluctuations in the OC produce a novel relationship between the fourth cumulant of the emitted charge and second cumulants (i.e., the variance or noise) of the emitted heat---a relation which explicitly renders the FCS non-Gaussian. This feature emerges both for equilibrium $T_{\rm in} =T_{\rm oc}$ and out-of-equilibrium $T_{\rm in} \neq T_{\rm oc}$ conditions. For negligible temperature fluctuations, the generally out-of-equilibrium output channel has Gaussian FCS. It is the finite OC heat capacity and the separation of time scales in the OC, Eq.~\eqref{eq:hierachy_time}, that produces a non-Gaussian FCS in the multi-edge channel-OC setup. 

\subsection{Third cumulant in the presence of finite voltage bias}
Until now, our analysis of the multi-edge-channel OC system focused on the case of no voltage bias in the system, i.e.,  $V_{\text{in},m} = 0$, so that there is no average potential on the OC either, $V_{\rm oc}=0$. Here, we will show that a finite $V_{\rm oc}$ is needed in order to obtain a finite correction, $\Delta\mathcal{C}_3$, to the odd cumulant $\mathcal{C}_3$ in Eq.~\eqref{eq:thid_cumulant}. This emergence of the third cumulant is of interest since detecting the third cumulant may be more experimentally feasible than the detection of the fourth cumulant, motivating the following additional calculation at finite $V_{\rm oc}$.

If we voltage bias a single input channel, (see Fig.~\ref{fig:Multi_Channel_Setup}), e.g., by taking $ V_{\text{in},m} = V \delta_{m,1}$, the energy dynamics of the system, given by Eq.~\eqref{eq:NChannel_heat_Dynamics}, is no longer fully described by heat currents. Instead we must in this equation take:
\begin{subequations}
\label{eq:JE_subst}
\begin{align}
    &J_{\text{in},m}(t)\rightarrow I^E_{\text{in},m}(t) = J_{\text{in},m}(t) - V \delta_{m,1} I_{\text{in},m}(t),\\
    & J_{\text{out},m}(t)\rightarrow I^E_{\text{out},m}(t) = J_{\text{out},m}(t) - V_{\rm oc} I_{\text{out},m}(t),
\end{align}
\end{subequations}
where from Kirchoff's voltage law, we have that $V_{\rm oc}=V/N$. From conservation of the average energy currents $\sum_m  I^E_{\text{in},m}  = \sum_m I^E_{\text{out},m} $, we find that the average OC temperature in the presence of the bias is fixed via
\begin{align} \label{eq:3C_temp}
       \kappa_m T_{\rm oc}^2  =   \kappa_0 T_{\rm in}^2 + \frac{N-1}{N^2}\frac{V^2}{ R_q},
\end{align}
where the thermal conductance $\kappa_m T_{\rm oc}$ is given in Eq.~\eqref{eq:J_out_m}. 

With the substitutions~\eqref{eq:JE_subst}, the energy conservation equation~\eqref{eq:energycons_Ncase} determining the OC temperature fluctuations becomes
\begin{align}\label{eq:energycons_3C}
    & \frac{d\Delta U (t)}{dt}\approx C_E[T_{\rm oc}] \frac{d\Delta T_{\rm oc}(t)}{dt} \notag  \\ &= \sum_{m=1}^N \Big[\delta J_{\text{in},m}(t) - \delta J^{\rm tot}_{\text{out},m}(t)  \notag \\ & -V  \left(\delta I_{\text{in},1}(t) \delta_{m,1} - N^{-1} \delta I_{\text{out},m}(t)\right)\Big],
\end{align} 
which thus contains a crucial term that is linear in the charge-current fluctuations. We now show that this modification results in a correction to the third cumulant of the emitted charge at finite bias. This correction, denoted $\Delta \mathcal{C}_3$, takes the form
\begin{align}
     &\Delta \mathcal{C}_3 \notag \equiv \prod_{n=1}^3 \int_{-\tau_{\rm obs}/2}^{\tau_{\rm obs}/2} dt_n\left\langle \left\langle \delta I(t_1) \delta I(t_2) \delta I(t_3) \right\rangle \right\rangle _E,
\end{align} 
where we remind that $\delta I(t)$ is short-hand notation for $\delta I_{\text{out},m}(t)$. Just as for the fourth cumulant, the third cumulant vanishes in the absence of temperature fluctuations due to the Gaussian nature of the charge fluctuations, leaving the correction $\Delta \mathcal{C}_3$ as the only non-vanishing component.   

With the same approach as for the fourth cumulant above, we write the leading order correction term as
\begin{align}
\label{eq:3C_corr}
    \Delta \mathcal{C}_3 &\approx  3\prod_{n=1}^2 \int_{-\tau_{\rm obs}/2}^{\tau_{\rm obs}/2} dt_n  \frac{\partial S_{\text{out},m}(0)
}{\partial T_{\rm oc}} \left\langle\langle \Delta T_{\rm oc}(t_1) \delta I_{}(t_2) \right\rangle\rangle_E.
\end{align} 
To compute these time integrals, we start by solving the energy conservation equation~\eqref{eq:energycons_3C} and then combine it with the heat Langevin equation~\eqref{eq:deltaJSol_Ncase}. We then apply the separation of fast and slow time scales according to Eq.~\eqref{eq:2p_approx}. Finally, by  using the identity for the charge-current noises at zero frequency~\eqref{eq:sumrule_Ncase_zerofreq}, we  arrive at 
\begin{align}
   &  \mathcal{C}_3 \approx \Delta \mathcal{C}_3 \notag \\ =
&3\tau_{\rm obs}  \left(\frac{N-1}{R_q N}\right)^2 \left(1- e^{-\tau_{\rm obs}/(2 \tilde{\tau}_{E,N})}\right)  \frac{ V  T_{\rm in}}{N \kappa_m T_{\rm oc} }. \label{eq:finite_TC}
\end{align}
This expression is indeed finite only when the voltage bias is finite $V\neq 0$ and for $N> 1$. The requirement of $N>1$ is, just as for the fourth cumulant, needed in order for the emitted charge current and zero-frequency charge current noise to depend on $T_{\rm oc}(t)$. 

The third cumulant~\eqref{eq:finite_TC} was obtained in Ref.~\cite{Pilgram_2003} for the particular case of $N=2$ and a symmetric cavity in the regime $\tau_{C,N} T_{\rm oc} \gg 1$. Our calculation thus generalizes the third cumulant to generic values of $\tau_{C,N}T_{\rm oc}$. We further note that for $N>2$, neither the third~\eqref{eq:finite_TC} nor the fourth cumulant~\eqref{eq:finite_FC} coincides with the corresponding cumulants in Ref.~\cite{Pilgram_2003} even in the regime $\tau_{C,N}T_{\rm oc}\gg 1$. This happens due to different definitions of the transmitted charge in the chiral and non-chiral conduction channels for $N>2$. We finally remark that a finite voltage bias would of course modify also the correction to the fourth cumulant, presented for the zero-voltage case in Eq.~\eqref{eq:finite_FC}. 

\section{Summary and outlook} 
We have presented a theory of charge and heat transport along quantum Hall edge channels strongly coupled to an Ohmic contact (OC), where the latter is characterized by not only a charge relaxation (or ``$RC$'') time, $\tau_C=R_q C$, but also an energy relaxation time $\tau_E\approx C_E[T_{\rm oc}]/(\kappa T_{\rm oc})$. We have thus comprehensively described the edge channel-OC system beyond the usually considered limit $\omega \tau_E\gg 1$. 

Based on the experimentally motivated time scale hierarchy in Eq.~\eqref{eq:hierachy_time}, we formulated a Langevin-based approach and determined the charge- and heat-current fluctuations emanating from the OC, as well as the OC potential and temperature fluctuations. Broadly, our findings reveal the delicate interplay between the charge and heat transport that is generated along  1D chiral edge channels when they are strongly coupled to an OC.

More specifically, we found that while only $\tau_C$ affects the charge transport, the heat transport is impacted by the combined effect of $\tau_C$ and $\tau_E$. Moreover, we established that the emitted charge-current noise, heat currents, and heat-current noise all display band-pass filtering effects due to the OC dynamics and highlight the out-of-equilibrium properties of the system. This is interesting from a out-of-equilibrium thermometry point-of-view and also extends the heat Coulomb blockade effect coming from the finite $RC$ time to other quantities and other time scales. Notably, the interplay of charge- and energy relaxation times results in a modified energy-relaxation time which governs the emitted heat current noise and generates a non-Gaussian full counting statistics of the OC's emitted charge. 

We end by discussing the feasibility to experimentally measure the predicted effects in this work. Experimental setups with 1D edge channels strongly coupled to an OC have been realized in both GaAs~\cite{Jezouin2013Nov} and in graphene~\cite{Srivastav2019Jul} devices. Moreover, charge currents and the charge-current noise are routinely measured in such devices and the charge-current noise dependence on the $RC$ time at low temperatures have also been observed, although so far only in GaAs~\cite{Sivre2018Feb}. To analyze also the heat-current fluctuations and the impact of $\tau_E$ will require two novel, important steps, which we believe should be within reach of state-of-the-art technology: (i) The fabrication of an OC, such that $\tau_E$ becomes sufficiently small. Either, this can be achieved with a sufficiently small-sized OC (see Sec.~\ref{sec:Setup_and_timescales} and Appendix~\ref{sec:Appendix_A} for estimates), or by exploiting the reduction of $\tau_E$ in a lower-dimensional OC. (ii) An experimental method to measure heat-current fluctuations in the QH regime. One possible way to measure these was proposed in Refs.~\cite{Furusaki1998Mar,Ebisu2022May}. In short,  quantum dots, side-coupled to the outgoing edge channels, can, via thermoelectricity, convert the edge channel heat-current fluctuations to more easily measurable charge-current fluctuations. Alternatively, heat-current fluctuations can be converted to temperature fluctuations via floating probe contacts~\cite{Dashti2018Aug}. Hence, complementing the setups in Figs.~\ref{fig:Single_Channel_Setup} and~\ref{fig:Multi_Channel_Setup} with quantum dots or probe contacts connected to the output channels extends the edge channel-OC system to a versatile platform to experimentally determine fundamental connections between heat and charge transport in a mesoscopic electronic circuit.

\begin{acknowledgments}
We thank Peter Samuelsson for fruitful discussions. We also acknowledge support from the Knut and Alice Wallenberg Foundation via the fellowship program and from the Swedish Vetenskapsrådet via Project No. 2018-05061 (J.S.) and Project No. 2023-04043 (C.S.), the Area of Advance Nano at Chalmers University of Technology (C.S and J.S.), and the Swiss National Science Foundation (F.S and E.S.). This project has received funding from the European Union's Horizon 2020 research and innovation programme under grant agreement No. 101031655 (TEAPOT). 
\end{acknowledgments}

\appendix
\section{Derivation of the OC heat capacity and the energy relaxation time}
\label{sec:Appendix_A}
In this Appendix, we compute the heat capacity, $C_E[T]$ and the energy relaxation time, $\tau_E$, under the assumption that the thermal properties of the OC can be described in terms of a free, three-dimensional (3D) Fermi gas with size $L^3$ and at temperature $T$. We further assume that the OC does not behave classically, i.e., we consider low temperatures $ T \ll T_\mathrm{F}$, where (in the units chosen here) the Fermi temperature is identical to the Fermi energy $E_\mathrm{F} = k_\mathrm{F}^2/(2m)$, with $k_\mathrm{F}$ the Fermi momentum and $m$ the electron mass.

In a free Fermi gas, the increase in internal energy due to a finite temperature $T$ compared to the zero-temperature limit, $\Delta U(T)$, is given as
\begin{align} \label{eq:Uchange}
	\Delta U(T) = \int_0^\infty \mathop{dE} E \mathcal{D}(E) f(E) - \int_0^{E_\mathrm{F}} \mathop{dE} E \mathcal{D}(E)\ .
\end{align}
Here, $f(E) = (e^{(E-\mu)/T}+1)^{-1}$ is the Fermi-Dirac distribution, $\mu$ is the electrochemical potential, which in the low-temperature case considered here equals the Fermi energy $E_\mathrm{F}$, and $\mathcal{D}(E)$ is the density of states. For a 3D gas, we have $\mathcal{D}(E) \propto \sqrt{E}$. Since the number of electrons, $\mathcal{N} = L^3k_\mathrm{F}^3/(3 \pi^2)$, is conserved when increasing the temperature from zero to $T$, we have the useful identity 
\begin{align}
\label{eq:useful}
    & \mathcal{N} = \int_0^{E_\mathrm{F}} \mathop{d E}  \mathcal{D}(E) = \int_0^{\infty} \mathop{d E}  \mathcal{D}(E) f(E)\notag \\
    & \Rightarrow	\int_0^{E_\mathrm{F}} \mathop{d E}  \mathcal{D}(E) (f(E)-1) + \int_{E_\mathrm{F}}^\infty \mathop{dE}  \mathcal{D}(E) f(E) = 0.
\end{align}
To compute the heat capacity $C_E[T]$, we differentiate Eq.~\eqref{eq:Uchange} with respect to the temperature and use Eq.~\eqref{eq:useful}. We then find
\begin{align}
\label{eq:CE_def}
	C_E[T]&\equiv \frac{d\Delta U(T)}{dT} = \int_{0}^\infty \mathop{dE} (E-E_\mathrm{F}) \mathcal{D}(E) \frac{\partial f(E)}{\partial T}\notag \\
 &\approx \mathcal{D}(E_\mathrm{F})  T \int_{-\infty}^\infty \mathop{dx}  \frac{x^2 e^x}{(e^x+1)^2} =\frac{\pi^2}{2}    \frac{T}{T_\mathrm{F}} \mathcal{N},
\end{align}
where we assumed that $T\ll T_\mathrm{F}=E_\mathrm{F}$  and take the value of the density of states at the Fermi energy $\mathcal{D}(E) \approx \mathcal{D}(E_\mathrm{F})\equiv 3 \mathcal{N}/(2 E_\mathrm{F})$.

We now compare $C_E[T]$ in Eq.~\eqref{eq:CE_def} to the Fermi gas level spacing, $\delta E$. The level spacing is given by the inverse density of states at the Fermi level $1/\mathcal{D}(E_\mathrm{F})$ and hence, for a $3$-dimensional Fermi gas of size $L^3$, 
\begin{align} \label{eq:Elvl_deriv_3D}
    \delta E\equiv\frac{1}{\mathcal{D}(E_\mathrm{F})}= \frac{2 E_\mathrm{F}}{3\mathcal{N}} = \frac{1}{(3\pi^2 \mathcal{N})^{\frac{1}{3}}} \frac{ \pi^2}{m L^2 },
\end{align}
which has the characteristic energy scale $\pi^2/(m L^2)$ and decreases with increasing electron number $\mathcal{N}$. Combining Eq.~\eqref{eq:Elvl_deriv_3D} with Eq.~\eqref{eq:CE_def}, gives the scaling $C_E[T] \sim    T \delta E^{-1}$ and the energy relaxation time thus increases linearly with the inverse level spacing:
\begin{align}
    \tau_E \equiv \frac{C_E[T]}{\kappa_0 T} \sim  \delta E_{}^{-1} ,
\end{align} where $\kappa_0=\pi/6$ is the heat conductance quantum.

For comparison, for a 2D Fermi gas, the level spacing is
\begin{align} \label{eq:Elvl_deriv_2D}
    \delta E_{2D}\equiv \frac{1}{\mathcal{D}_{2D}(E_\mathrm{F})}=\frac{E_\mathrm{F}}{\mathcal{N}_{2D}} = \frac{1}{\pi} \frac{ \pi^2}{m L^2 }, 
\end{align} with $\mathcal{D}_{2D}(E_\mathrm{F}) = E_\mathrm{F}/\mathcal{N}_{2D}$ and $\mathcal{N}_{2D} = k_\mathrm{F}^2 L^2/2\pi$. Note that the 2D gas level spacing~\eqref{eq:Elvl_deriv_2D} is independent of the 2D electron number $\mathcal{N}_{2D}$. 

Instead, for a 1D Fermi gas we find
\begin{align} \label{eq:Elvl_deriv_1D}
    \delta E_{1D}\equiv \frac{1}{\mathcal{D}_{1D}(E_\mathrm{F})} = \frac{E_\mathrm{F}}{2\mathcal{N}_{1D}} = \frac{\mathcal{N}_{1D}}{4} \frac{ \pi^2}{m L^2 }, 
\end{align}  with $\mathcal{D}_{1D}(E_\mathrm{F}) = E_\mathrm{F}/(2\mathcal{N}_{1D})$ and $\mathcal{N}_{1D} = k_\mathrm{F} L/\pi$.  Hence,  the 1D gas level spacing increases linearly with the 1D electron number $\mathcal{N}_{1D}$.

\section{Bosonized model of the Ohmic contact and derivation of the heat-charge current relation}
\label{sec:Appendix_B}
With the bosonization technique (see e.g., Refs.~\cite{vonDelft1998Nov,slobodeniuk_equilibration_2013,Sukhorukov2016Mar}), the strongly coupled edge channel-OC system can be treated exactly. The bosonized Hamiltonian of the edge-OC system is given by~\cite{matveev_coulomb_1995,furusaki_theory_1995,slobodeniuk_equilibration_2013}
\begin{align}
\label{eq:Bos_Ham}
    \mathcal{H} = \frac{v_\mathrm{F}}{4 \pi} \sum_{i=\mathrm{in,out}}\int_{-\infty}^\infty \mathop{dx} \left(\partial_x \phi_i(x,t)\right)^2 + \frac{Q(t)^2}{2C},
\end{align}
where $i\in \{\text{in},\text{out}\}$ labels the incoming and outgoing edge channels, $C$ is the OC charge capacitance, and $Q(t) = \sum_{i'} \int_{-\infty}^0 \mathop{dx} \rho_{i'}(x,t) \exp(\epsilon x/v_\mathrm{F})$ is the OC charge operator. Here, the charge is expressed as integrated charge densities $\rho_i(x,t)\equiv\frac{e}{2\pi}\partial_x \phi_i(x,t)$ inside the OC region $x\in (-\infty,0]$. The exponentially decaying term arises due to the finite lifetime of excitations $\epsilon^{-1}>0$ inside the OC. Importantly, the Hamiltonian~\eqref{eq:Bos_Ham} is quadratic in the bosonic fields and can thus be solved exactly. The equations of motion for the bosons are obtained from the Heisenberg equation $\partial_t \phi_i(x,t)=-i\left[\phi_i(x,t),\mathcal{H}\right]$ and are supplemented by the boundary conditions $\partial_t\phi_{\rm in}(+\infty)=-2\pi I_{\rm in}(t)/e$ and $\partial_t\phi_{\rm out}(-\infty)=-2\pi I_{\rm oc}(t)/e$. The solution of the equations of motion can be re-arranged precisely as the charge dynamics in Eqs.~\eqref{eq:Kirchhoff} and~\eqref{eq:Langevin}~\cite{slobodeniuk_equilibration_2013}.

Furthermore, by inspecting the first term in Eq.~\eqref{eq:Bos_Ham}, we obtain the heat current operator, $J$, for a single channel. In 1D, this operator must have the form ``velocity$\times$energy density'', for which the only consistent choice is 
\begin{align}
    J(x,t)=\frac{v_\mathrm{F}^2}{4\pi}\left(\partial_x \phi_i(x,t)\right)^2.
\end{align}
Identifying in this expression the charge-current operator $I(x,t)=v_\mathrm{F}\rho(x,t) = e v_\mathrm{F} \partial_x \phi(x,t)/(2\pi)$ and the resistance quantum $R_q=2\pi/e^2$, leads directly to the heat-charge current relation in Eq.~\eqref{eq:JI_relation}. 

\section{Equivalence of fermionic and bosonic descriptions of the average heat current}
\label{sec:Appendix_equivalence}
For completeness, we provide in this Appendix a derivation (previously presented in Ref.~\cite{levkivskyi_energy_2012}) of the equivalence between: (i) The equilibrium heat current~\eqref{eq:JT_rel}, given in terms of bosonic density fluctuations, and (ii) A description of the equilibrium heat current in terms of an electronic distribution function of chiral fermions. The latter quantity
is given by
\begin{align}\label{eq:Jfermionic}
    J_{\rm ferm} = \frac{1}{2\pi}\int_{-\infty}^{\infty}d\epsilon (\epsilon-\mu)\left[ f(\epsilon-\mu)-\theta(-\epsilon+\mu) \right] = \frac{\pi T^2}{12},
\end{align} 
where we explicitly included the electrochemical potential $\mu$ for clarity. The definition of the fermionic distribution function $f(\epsilon)$ is given as
\begin{align}
    & f(\epsilon) \equiv \int_{-\infty}^{\infty}dt e^{i \epsilon t} K(t),\label{eq:ferm_corr_K}\\
    & K(t)=\left\langle \psi^\dagger(t) \psi(0) \right\rangle,\label{eq:ferm_corr}
\end{align}
where $\psi^\dagger(t)$ is a fermionic operator creating a fermion at time $t$. In the time-domain, we express the average fermionic heat current~\eqref{eq:Jfermionic}, via Eqs.~\eqref{eq:ferm_corr_K} and~\eqref{eq:ferm_corr}, as
\begin{align}
\label{eq:J_ferm_time}
     J_{\rm ferm} = -i \left.\partial_t \left[K(t)-\frac{1}{2\pi i t}\right]\right|_{t\rightarrow 0},
\end{align}
where the second term in the brackets is the subtracted zero-temperature contribution. Next, we express the fermionic correlation function~\eqref{eq:ferm_corr} with the bosonization technique~\cite{Sukhorukov2016Mar} as
\begin{align}
    K(t) = \frac{1}{2\pi a}\left\langle e^{-i \phi(t)} e^{i \phi(0)} \right\rangle = \frac{1}{2 \pi a} e^{\left\langle (\phi(t)- \phi(0))\phi(0)\right\rangle},\label{eq:K_bosonic}
\end{align}
where $\phi(t)$ is a chiral bosonic field and $a$ is a short-distance cutoff. The last equality in Eq.~\eqref{eq:K_bosonic}, i.e., the formulation of the fermionic correlation function in terms of a bosonic correlation function, holds whenever the bosonic theory is quadratic, or ``Gaussian'', which is particularly true in equilibrium. Furthermore, the boson $\phi(t)$ is related to the current operator as
\begin{align}
\label{eq:I_and_boson}
    I(t) = -e\frac{\partial_t \phi(t)}{2\pi},
\end{align}
which permits us to express the bosonic correlation function in terms of the equilibrium charge-current noise as
\begin{align}
\label{eq:boson_noise_relation}
    -\frac{e^2}{(2\pi)^2}\omega \omega'\langle \phi(\omega)\phi(\omega') \rangle &= \langle \delta I(\omega)\delta I(\omega')\rangle \notag \\ & \equiv 2\pi \delta(\omega+\omega') S(\omega),
\end{align}
where we assumed $\langle I(\omega) \rangle = \langle \phi (\omega) \rangle = 0$. Using Eq.~\eqref{eq:boson_noise_relation} in Eq.~\eqref{eq:K_bosonic}, we write the fermionic correlation function in terms of the charge-current noise as
\begin{align}
     \ln K(t) = -\ln(2\pi a)+  \frac{2\pi}{e^2}\int_{-\infty}^{\infty}\frac{\mathop{d\omega}}{\omega^2} S(\omega) (e^{-i\omega t}-1).
\end{align}
Next, we identify $R_q=2\pi/e^2$ and add and subtract the term $\ln (t)$, which further allows us to write
\begin{align}
    &\ln K(t) = -\ln(2\pi a) + \ln K_n(t) - \ln(t),\\
&\ln K_n(t)= R_q\int_{-\infty}^{\infty}\frac{\mathop{d\omega}}{\omega^2} \left(S(\omega)-\frac{\omega\theta(\omega)}{R_q}\right) (e^{-i\omega t}-1).
\end{align}
By expanding $\ln K(t)$ to second order in $t$, dropping odd terms in the integrand, and inserting into Eq.~\eqref{eq:J_ferm_time}, we finally arrive at
\begin{align}
\label{eq:Jferm_as_noise}
 J_{\rm ferm} &= \left.-\frac{1}{4\pi} \partial_t^2 \ln K_n(t)\right|_{t\rightarrow0 } \notag\\  &=  \frac{R_q}{4\pi}\int_{-\infty}^\infty d\omega \left[ S(\omega)-\frac{\omega\theta(\omega)}{R_q}\right],
\end{align}
which is precisely the equilibrium heat current in terms of the charge-current noise, as presented in Eq.~\eqref{eq:JT_rel}.

All the above steps can in fact be repeated verbatim for the output heat current. It is defined in terms of an output, generally out-of-equilibrium, distribution function $f_{\rm out}(\epsilon)$ as
\begin{align}
\label{eq:J_out_App}
  J_{\rm out} =  \frac{1}{2\pi}\int_{-\infty}^{\infty}d\epsilon (\epsilon-\mu)\left[ f_{\rm out}(\epsilon)-\theta(-\epsilon+\mu) \right],
\end{align}
which leads to Eq.~\eqref{eq:Jferm_as_noise} with the replacement $S(\omega)\rightarrow S_{\rm out}(\omega)$. By inspecting the output current~\eqref{eq:jSol} and the boson-current identity~\eqref{eq:I_and_boson}, we see that the output boson must be a linear combination of the input and OC bosons. This implies further that the output charge-current noise is a linear combination of the input and OC charge-current noises, as shown in Eq.~\eqref{eq:Sout}. Thus, it follows that the output heat current~\eqref{eq:J_out_App} is exactly equivalent to the formulation in Eq.~\eqref{eq:T_integral}.

%

\end{document}